\documentclass[a4paper,12pt,oneside]{article}
\usepackage[cp1251]{inputenc}
\usepackage[english]{babel}
\usepackage{graphicx}
\usepackage{epsfig}
\usepackage{enumerate}
\usepackage{amsmath,amsfonts,amsthm}
\usepackage{amssymb}
\usepackage{fullpage}
\usepackage{cite}

\title{THERMAL STRUCTURE OF A PROTOSTELLAR ENVELOPE}

\author{Ya.N. Pavlyuchenkov$^1 \footnote {E-mail: pavyar@inasan.ru},$ A.G. Zhilkin$^{1,2}$, E.I. Vorobyov$^{3,4}$, \\
A.M. Fateeva$^1$}

\date{
$^1$ Institute of Astronomy, Russian Academy of Sciences,
Pyatnitskaya st. 48, Moscow, 119017 Russia \\
$^2$ Chelyabinsk State University, Chelyabinsk, Russia \\
$^3$ Research Institute of Physics, Southern Federal University, Rostov-on-Don 344090, Russia \\
$^4$ Institute of Astrophysics, University of Vienna, Vienna,
Austria }

\begin{document}

\maketitle

\abstract{A numerical hydrodynamical model for the evolution of
spherically symmetric collapsing clouds, designed for the
calculation of the thermal structure of these objects in both the
prestellar and protostellar stages of their evolution, is
presented. Distinctive features of the model include the
possibility of independently describing the temperatures of the
gas and dust, which is extremely important when calculating the
thermal structure of prestellar and protostellar clouds, and the
account of the radiation flux from the central protostar. This
model is used to compare the theoretical density and temperature
distributions with observations for nearby sites of star formation
obtained with the Herschel Space Observatory. Application of the
diffusion approximation with a flux limiter describes well the
radial density and temperature distributions in protostellar
clouds. However, significant differences between the model and
observational density profiles were found for prestellar stages,
suggesting the presence of appreciable deviations from equilibrium
in the prestellar clouds. An approximate method for calculating
the thermal structure of a cloud based on the adaptive
$\tau$-approximation is presented. Application of the
$\tau$-approximation yields good agreement with the diffusion
approximation for the prestellar phase, but produces appreciable
discrepancies for the protostellar phase, when the thermal
structure of the accreting envelope is determined by the radiation
of the protostar.}

\section{Introduction}

A model for calculating the thermal structure of a collapsing
protostellar cloud was presented in ~[1, 2]. This model is fast
enough for use in hydrodynamic calculations, and is also
sufficiently accurate to use the modeling results to calculate
theoretical spectra that can be compared with observations. The
key feature of the model is separate treatment of dust and gas
temperatures, since these two temperatures may differ
significantly in early phases of the evolution of the clouds, and
also in their outer regions. Different heating and cooling
mechanisms are considered for the gas and dust, while they
exchange energy through collisions. Since the main source of
heating (cooling) of the dust is the absorption (emission) of
radiation, the method used to calculate the radiation transfer is
an important component of the model.

The basic idea of the radiative transfer computation method is to
divide the overall frequency range into low-frequency (infrared)
and high frequency (ultraviolet) parts. The high-frequency part is
diluted by interstellar radiation. The intrinsic radiation of the
dust can be neglected in this part of the spectrum, and the
absorption (and scattering) of external radiation alone can be
considered. In the low-frequency part of the spectrum, it is
necessary to consider both absorption and thermal emission of the
ambient medium. This division of the spectral range into two parts
enables the use of different appropriate approximate methods in
each. Modeling the UV part of the spectrum reduces to calculating
the mean intensity of the UV radiation via direct integration of
the radiative transfer equation along selected directions. The
calculation of the intensity of the IR radiation is based on
solving a system of moment equations that represent the diffusion
approximation. Thus, the model contains four interacting
components: gas, dust, and the IR and UV radiation.

An axially symmetric realization of this thermal model was used to
calculate the early evolution of a contracting, magnetized
protostellar cloud [1]. Modeling of later stages, up to the
formation of the first hydrostatic core, was performed in a
spherically symmetric approximation in [2]. In our present study,
we describe a modification of the model designed for the
calculation of the accretion phase in the evolution of a
protostellar cloud following the formation of the protostar. We
present the results of our calculations of the structure of a
spherically symmetric, contracting protostellar cloud and
accreting envelope. To simplify the computation of the thermal
structure, we propose the adaptive $\tau$-approximation method and
compare it with the original method. We also compare the model and
observed distributions of the density and temperature in
prestellar and protostellar cores of molecular clouds.

\section{MODIFICATION OF THE MODEL}

Pavlyuchenkov and Zhilkin [2] investigated the thermal evolution
of a protostellar cloud up to the formation of the first
hydrostatic core. Evolution of this core subsequently leads to the
evaporation of dust, dissociation of hydrogen, and, consequently,
to the collapse of the first core and the formation of a second
hydrostatic core, which is essentially the protostar [3–5].
Simulation of these stages requires that additional physical
processes be taken into account in the model for the protostellar
cloud. The timescales for these stages are extremely short
compared to the dynamical timescale of the accreting envelope.
This makes hydrostatic modeling of the core and accreting envelope
using a single model extremely difficult. Moreover,
multi-dimensional models must be used in studies of hydrostatic
cores, since their formation and evolution are closely associated
with the formation of accretion (protoplanetary) disks and
outflows from them. However, in addition to modeling the evolution
of a protostar and its surroundings (the protostellar disk),
studies of the structure of the extended accreting envelope
(including its thermal structure) are of interest. Accretion from
the envelope supplies matter to the protostar and protostellar
disk, determining their evolution. In turn, the observational
identification of protostellar accreting cores depends on
assumptions about their thermal structure, which requires
appropriate models.

Historically, the simplest models for prestellar and protostellar
envelopes were based on semianalytical self-similar solutions of
the hydrodynamical equations in isothermal or polytropic
approximations~[7, 8], with possible modifications to take into
account rotation in the cloud~[9], as well as numerical
hydrodynamical modeling taking into account the finite size of the
cloud~[10–12]. Later, numerical models were used to consider the
nonisothermal structure of the envelopes more accurately, via more
accurate solution of the transfer equation taking into account the
frequency dependence of the absorption coefficient~[13], and
applying the diffusion approximation with a flux limiter and
either mean~[14] or frequency-dependent~[15] opacities. Modeling
of protostars and protostellar envelopes has been carried out by a
number of research groups (see, e.g.,~[6]). However, most models
still assume that the temperatures of the gas and dust are equal,
since frequent collisions should establish thermal equilibrium
between these components. As is noted above, however, this
assumption may not be valid at some stages of the evolution. In
some models, the dust and gas were treated separately by
postprocessing the gas density distribution in the envelope using
Monte-Carlo methods for the radiative transfer~[16]. Finally, most
studies have focused on early stages in the evolution of the first
hydrostatic core and protostar~[17]. Here, we consider a
collapsing cloud and describe a modification of our model~[2]
designed for the calculation of the thermal structure of an
accreting envelope.

The model of~[2] is based on a numerical solution of the
hydrodynamical and radiative transfer equations. The continuity
equation, equation of motion, and the equation describing the
variations of the thermal energy of the gas $E_{\textrm{g}}$ have
the form

\begin{eqnarray}
\dfrac{\partial \rho}{\partial t} + \nabla \cdot \left(\rho
{\mathbf{v}}\right) = 0,
\label{hd01}\\
\dfrac{\partial {\mathbf{v}}}{\partial t} + \left({\mathbf{v}}
\cdot \nabla\right) {\mathbf{v}} = {-}\dfrac{\nabla p}{\rho} +
{\mathbf{g}},
\label{hd02}\\
\dfrac{\partial E_{\textrm{g}}}{\partial t} + \nabla \cdot \left(
E_{\textrm{g}} {\mathbf{v}} \right) + p  \nabla \cdot {\mathbf{v}}
= \Gamma_{\textrm{g}} - \Lambda_{\textrm{g}} -
\Lambda_\textrm{gd}. \label{hd03}
\end{eqnarray}

When the velocity of the medium is low, the equations for the
thermal energy of the dust $E_{\textrm{d}}$ and the energy of the
IR radiation $E_{\textrm{r}}$ have the form

\begin{eqnarray}
\dfrac{\partial E_{\textrm{d}}}{\partial t} =
-\sigma_P(aT_{\textrm{d}}^4 - E_{\textrm{r}}) +
\Lambda_\textrm{gd} + S,
\label{rtmov01}\\
\dfrac{\partial E_{\textrm{r}}}{\partial t} =  \nabla \cdot
\left(\dfrac{1}{\sigma_R}\nabla E_{\textrm{r}}\right) +
\sigma_P(aT_{\textrm{d}}^4 - E_{\textrm{r}}). \label{rtmov02}
\end{eqnarray}

In these equations, $\rho$ is the density, ${\mathbf{v}}$ the
velocity, $p$ the pressure, $\mathbf{g}$ the gravitational
acceleration, $\Gamma_{\textrm{g}}$ the gas heating function,
$\Lambda_{\textrm{g}}$ the gas cooling function,
$\Lambda_\textrm{gd}$ the rate of exchange of thermal energy
between the gas and dust, $S$ the rate of dust heating by UV
radiation, $T_{\textrm{d}}$ the dust temperature, $\sigma_R$ and
$\sigma_P$ coefficients that depend on the opacity of the medium,
and a the radiative constant. A detailed description of the system
of equation and the solution method is given in~[2]. Here, we will
discuss our modification of this model.

The modification was carried out in the framework of the
well-known ``sink cell'' formalism (see, e.g.,~[18]). In this
approach, the central region of the cloud (the sink cell) is
excluded from the simulation when a high density is achieved
there. This makes it possible to exclude from the modeling the
rapid evolution of this region with numerous physical processes
leading to the formation of the central star. However, the
influence of this region on the surrounding cloud is taken into
account, as well as the influence of the cloud on the inner
region. This interrelation has both a dynamical and thermal
nature. Thus, the modification of the model affects the
hydrodynamical method and the method used to compute the thermal
structure. The modification of the hydrodynamical method depends
on the type of method used (Eulerian, Lagrangian, smoothed
particle hydrodynamics method, etc.).

We will now describe the changes in the one-dimensional Lagrangian
method used. Let $r_{\textrm{cell}}$ be the radius of the inner
sink cell. As the cloud contracts, the cells of the computational
domain become denser and move toward the center. We will suppose
that a cell is excluded from the calculations (it is absorbed by
the sink cell) as soon as the right-hand border of the cell
crosses the radius $r_{\textrm{cell}}$. The next cell then becomes
the boundary cell, and the cells are renumbered such that the new
boundary cell is assigned the subscript ``1''. Thus, the left-hand
boundary of the new boundary cell $r_{1}$ will be located to the
left of $r_{\textrm{cell}}$, while its right-hand boundary $r_{2}$
is located to the right of $r_{\textrm{cell}}$. We assume that the
pressure gradient vanishes at the left border ($r_{1}$):

\begin{eqnarray}
\left.\dfrac{dp}{dr}\right|_{r_{1}}=0.
\end{eqnarray}

This condition means that the matter flows freely (by inertia)
into the inner zone under the action of gravity of the sink cell
alone.

The modification of the method used to compute the thermal
structure involves two operations: taking into account the sink
cell and modifying the diffusion operator. Taking into account the
sink cell reduces to formulating a new boundary condition at the
inner boundary of the computational domain $r_{1}$, where the IR
flux from the sink cell $L_{\textrm{cell}}$ is determined by the
sum of the photospheric and accretion luminosities of the young
star:

\begin{eqnarray}
L_{\textrm{cell}}  = L_{\textrm{star}} + L_{\textrm{acc}}.
\end{eqnarray}

We calculated the photospheric luminosity using the stellar model
of~[19, 20], where the luminosity $L_{\textrm{star}}$ and the
radius of the young star $R_{\textrm {star}}$ are computed as
functions of the instantaneous $M_{\textrm{star}}$ and age
$t_{\textrm{star}}$. Assuming equipartition between the thermal
and kinetic (mechanical) energy of the accreting
matter\footnote{The mechanical energy does not contribute to the
    radiation and is mostly carried away together with polar jets and
    outflows of matter.},  the
accretion luminosity is given by

\begin{eqnarray}
L_{\textrm{acc}} =
\frac{1}{2}\dfrac{G\dot{M}M_{\textrm{star}}}{R_{\textrm{star}}},
\end{eqnarray}
where
\begin{eqnarray}
\dot{M} = 4\pi r_1^2 \rho_{1} v_{1}
\end{eqnarray}

is the accretion rate through the inner boundary of the
computational domain. Here, $\rho_{1}$ is the density in the first
(boundary) cell, and $v_{1}$ is the radial velocity at the
left-hand border of the first cell.

When calculating the accretion luminosity using these formulas, we
assumed that all the matter crossing the boundary of the
computational domain is instantaneously accreted onto the star,
and the all thermal energy of the accreted matter is converted
into IR radiation; this corresponds to the so-called cold
accretion limit, and is valid if the accretion rate onto the
protostar is not too high, $\dot{M} \lesssim
10^{-5}~M_{\odot}$/yr~[21]. Since the coordinate in the Lagrangian
method is the mass inside the corresponding radius, the mass
inside the absorbing zone can be calculated directly from the
coordinates of the boundary cell:

\begin{eqnarray}
M_{\textrm{star}} = 4\pi q_{1},
\end{eqnarray}
where $q_{1}$ is mass (Lagrangian) coordinate.

Finite-difference approximation of the diffusion operator in the
 spherically-symmetric case\\

$\nabla \cdot \left(\dfrac{1}{\sigma}\nabla E\right) =
\dfrac{1}{r^2}\dfrac{\partial}{\partial{r}}
\left(\dfrac{r^2}{\sigma}\dfrac{\partial E}{\partial r}\right)$
has the form\footnote{Note that there is an error in Eq. (38) in
[2].}:

\begin{eqnarray}
\hat{\Lambda} E = \dfrac{1}{R_{i+1/2}^2 \Delta r_{i+1/2}}\left(
\dfrac{R_{i+1}^2}{\sigma_{i+1}}\dfrac{E_{i+3/2}-E_{i+1/2}}{\Delta
r_{i+1}}-
\dfrac{R_{i}^2}{\sigma_{i}}\dfrac{E_{i+1/2}-E_{i-1/2}}{\Delta
r_{i}} \right),
\end{eqnarray}
where
\begin{eqnarray}
R_{i+1/2}^2 = \dfrac{1}{3}(r_{i}^2+r_{i}r_{i+1}+r_{i+1}^2), \\
R_{i}^2=r_{i-1/2}r_{i+1/2}, \\
\Delta r_{i} = r_{i+1/2}-r_{i-1/2},\\
\Delta r_{i+1/2} = r_{i+1}-r_{i}.
\end{eqnarray}

Taking into account the new boundary condition, the
finite-difference approximation of this operator for the inner
boundary of the cell takes the form

\begin{eqnarray}
\hat{\Lambda} E = \dfrac{1}{R_{1/2}^2 \Delta r_{1/2}}\left(
\dfrac{R_{1}^2}{\sigma_{1}}\dfrac{E_{3/2}-E_{1/2}}{\Delta r_{1}}+
\dfrac{L_\text{cell}}{4\pi} \right),
\end{eqnarray}

We introduce the sink cell when the central density becomes
$n(\textrm{H}_2)=10^{14}$~cm$^{-3}$. The central temperature at
this time is close to  $1000$~K. The radius of the sink cell we
used is $r_{\textrm{cell}}=50$~AU. This is appreciably larger than
the radius of the first hydrostatic core, but appreciably smaller
than the extent of the accreting envelope. This choice ensures
that the sink cell is larger than the first hydrostatic core.
Otherwise, the formation of a quasi-hydrostatic structure at the
inner boundary of the computational mesh would hinder the
accretion of matter onto the sink cell. When the sink cell is
introduced, all the grid cells inside this zone are eliminated
from further consideration. The time when the sink cell is
introduced is also taken to be the time when the star forms, when
we start to compute the photospheric and accretion luminosity.

Other numerical criteria for the formation of the protostar are
also possible. In particular, it was found in~[4] that the maximum
mass of the first hydrostatic core (i.e., the mass at the
beginning of the dissociation of molecular hydrogen and the
subsequent formation of the protostar) was close to
$0.05~M_{\odot}$, independent of the initial mass of the cloud.
This can be used to define the moment of the protostar’s
formation. However, this value was obtained for nonrotating,
spherically symmetric clouds. Taking into account the angular
momentum of the cloud could lead to a reassessment of the maximum
mass of the first hydrostatic core~[22].

The Eddington approximation was applied in the equation for the
energy of the IR radiation in~[2]. This approximation assumes that
the radiation field is close to isotropic, making it possible to
express the radiation pressure tensor $\hat{P_{\textrm{r}}}$ in
terms of the radiation energy $E_{\textrm{r}}$ in the form
$\hat{P_{\textrm{r}}}=\dfrac{1}{3}\hat{I}E_{\textrm{r}}$, where
$\hat{I}$ is the unit tensor. The diffusion coefficient in~(5) is
then

\begin{eqnarray}
D_{\textrm{E}}=1/\sigma_R = \dfrac{c}{3\rho_{\textrm{d}}
    \kappa_R},
\end{eqnarray}

where $\rho_{\textrm{d}}$ is the dust density, $\kappa_R$ the
Rosseland mean opacity, and $c$ the speed of light. However, the
Eddington approximation yields large errors when the radiation
propagates from a localized source in an optically thin medium (in
the so-called streaming regime). The accretion stage in the
evolution of a protostellar cloud proceeds in precisely this
regime, with the protostar playing the role of the localized
source and the accreting envelope being optically thin in the IR.
Therefore, in place of the diffusion coefficient in the Eddington
approximation, we used a flux limiter (so-called Flux Limited
Diffusion, FLD)~[23].

If the frequency-averaged scattering coefficient is substantially
lower than the absorption coefficient (as is true in the IR), the
diffusion coefficient is given by

\begin{eqnarray}
D_{\textrm{FLD}} = \lambda\dfrac{c}{\rho_{\textrm{d}} \kappa_P}
\dfrac{aT_{\textrm{d}}^4}{E_{\textrm{r}}}, \label{FLD}
\end{eqnarray}
where
\begin{eqnarray}
\lambda = \dfrac{2+f}{6+3f+f^2}, \quad {\mathbf{f}} ={-}
\dfrac{\nabla E_{\textrm{r}}}{\rho_{\textrm{d}}\kappa_P
    aT_{\textrm{d}}^4}, \label{R_FLD}
\end{eqnarray}

where $\kappa_P$ is the Planck mean absorption coefficient. In the
iterative scheme used to derive $T_{\textrm{d}}^{n+1}$ and
$E_{\textrm{r}}^{n+1}$, the values of $D_{\textrm{FLD}}$ were
computed using the dust temperature $T_{\textrm{d}}^{n}$ d and the
energy of the IR radiation $E_{\textrm{r}}^{n}$ from the previous
time step. The FLD approximation operates well in both the
streaming regime and the optically thick regime, when
$D_{\textrm{FLD}} \approx D_{\textrm{E}}$. Thus, the evolution of
the accreting envelope and the preceding stage of the formation of
the hydrostatic core should both be computed correctly in the FLD
approximation.

\begin{figure*}[t!]
    \begin{center}
    \includegraphics[scale=0.7]{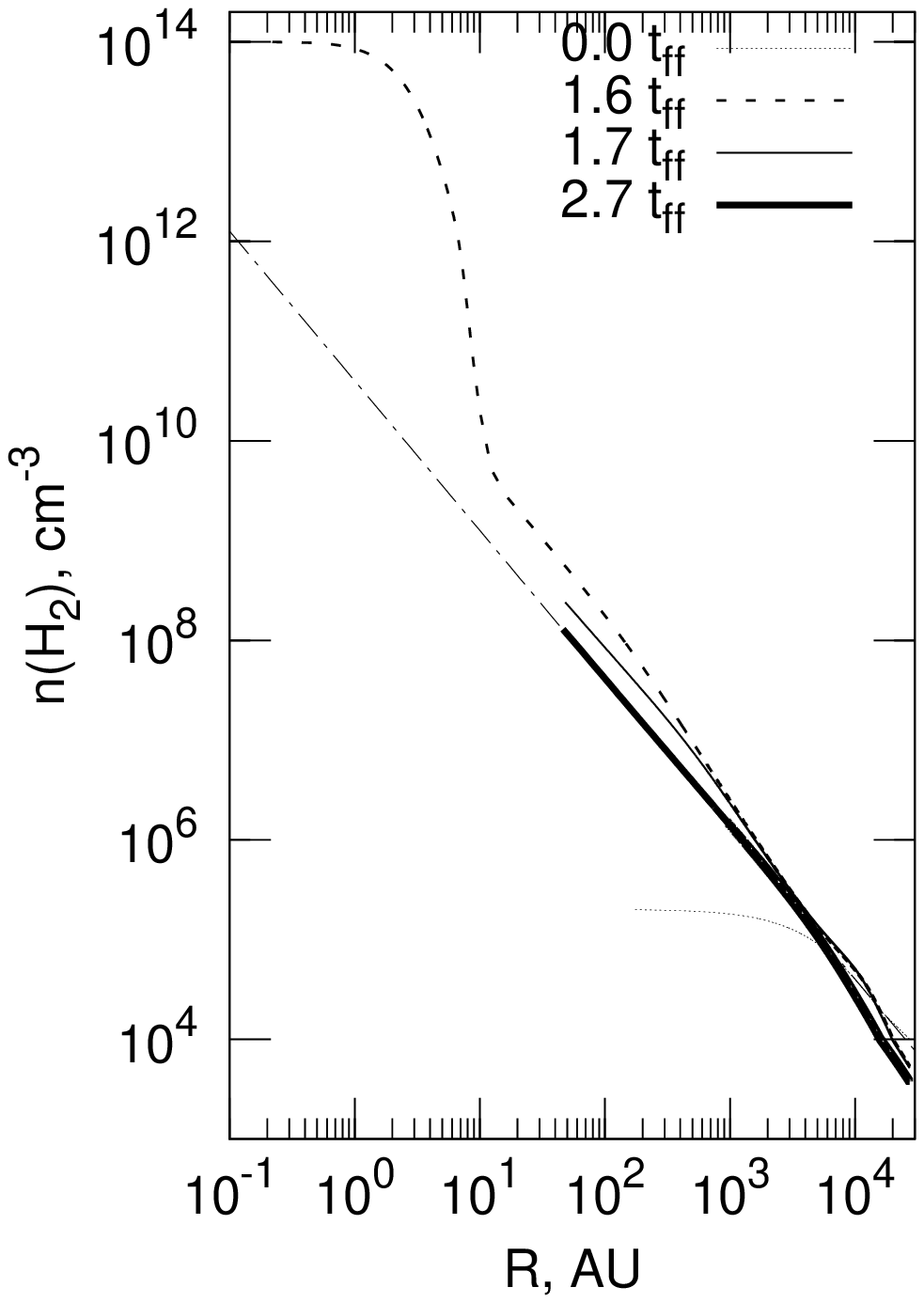}
    \includegraphics[scale=0.7]{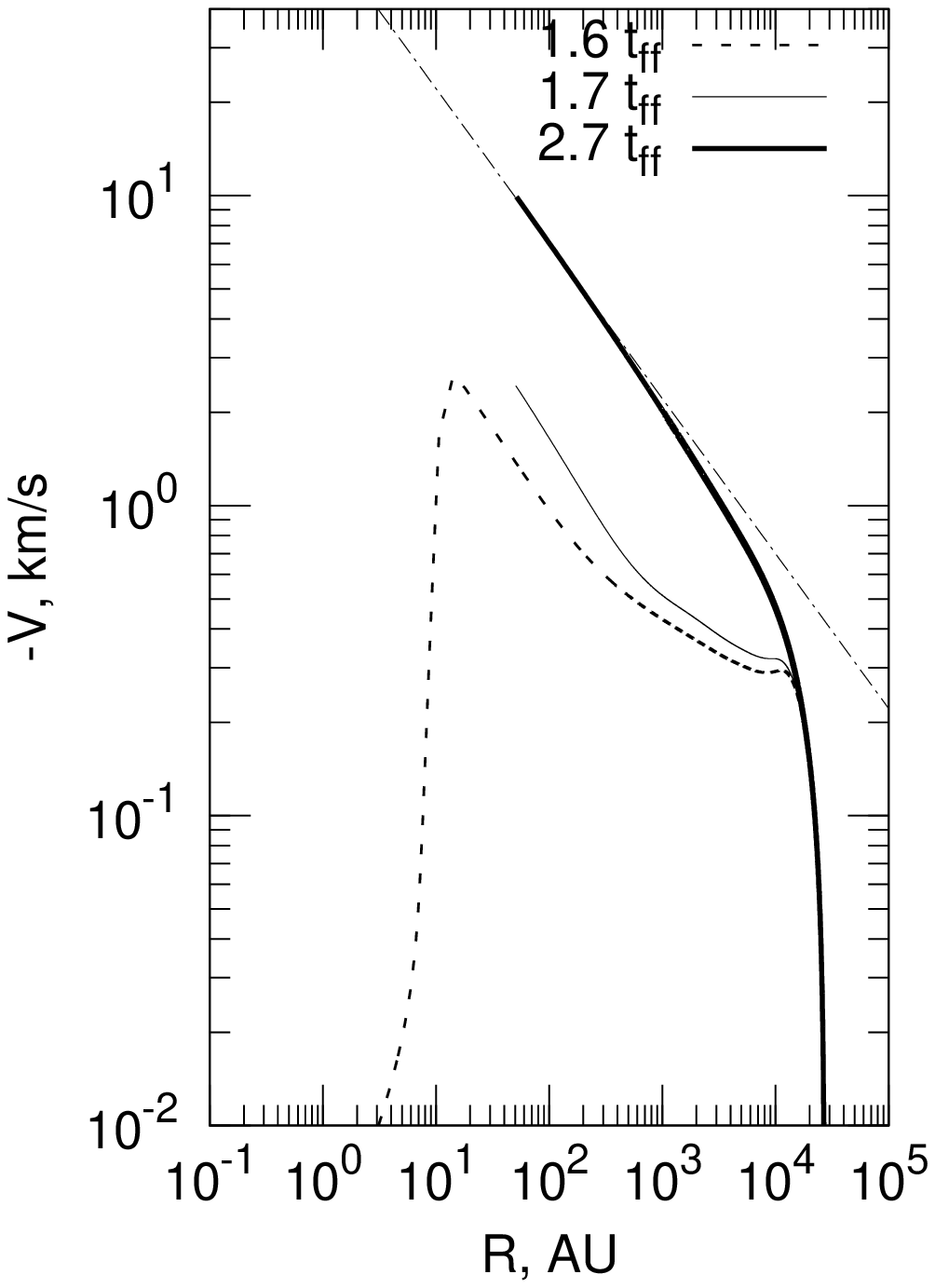}
    \end{center}
    \caption{Distributions of the hydrogen number density (left) and the
velocity (right) in the protostellar cloud at various times during
its evolution. The dash.dot lines show the functions $n \propto
r^{-3/2}$ and $v \propto r^{-1/2}$, whose indices are defined by
the self-similar solution in the case of spherically symmetric
accretion.
 \hfill}
\end{figure*}

\begin{figure*}[t!]
    \includegraphics[scale=0.7]{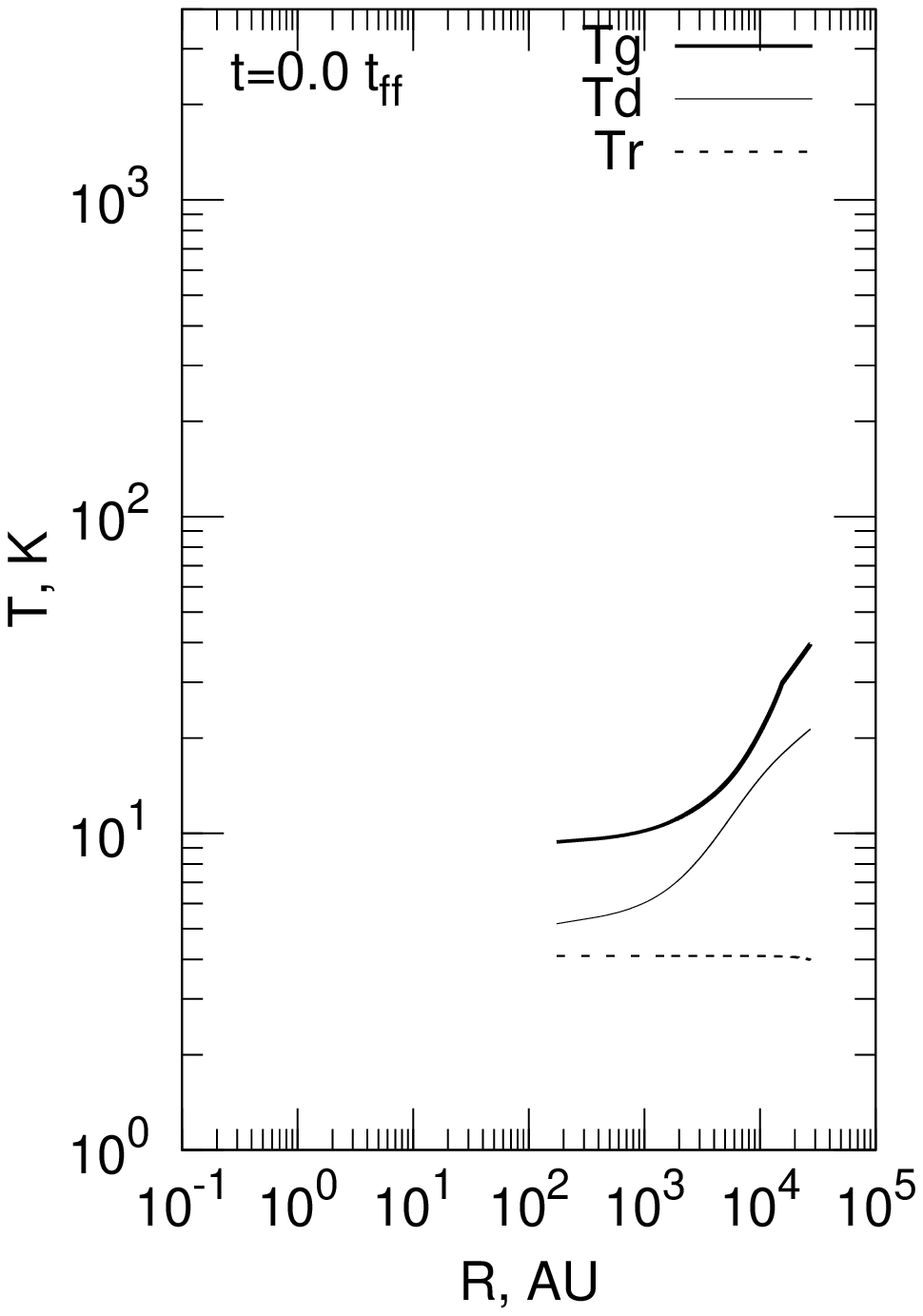}
    \includegraphics[scale=0.7]{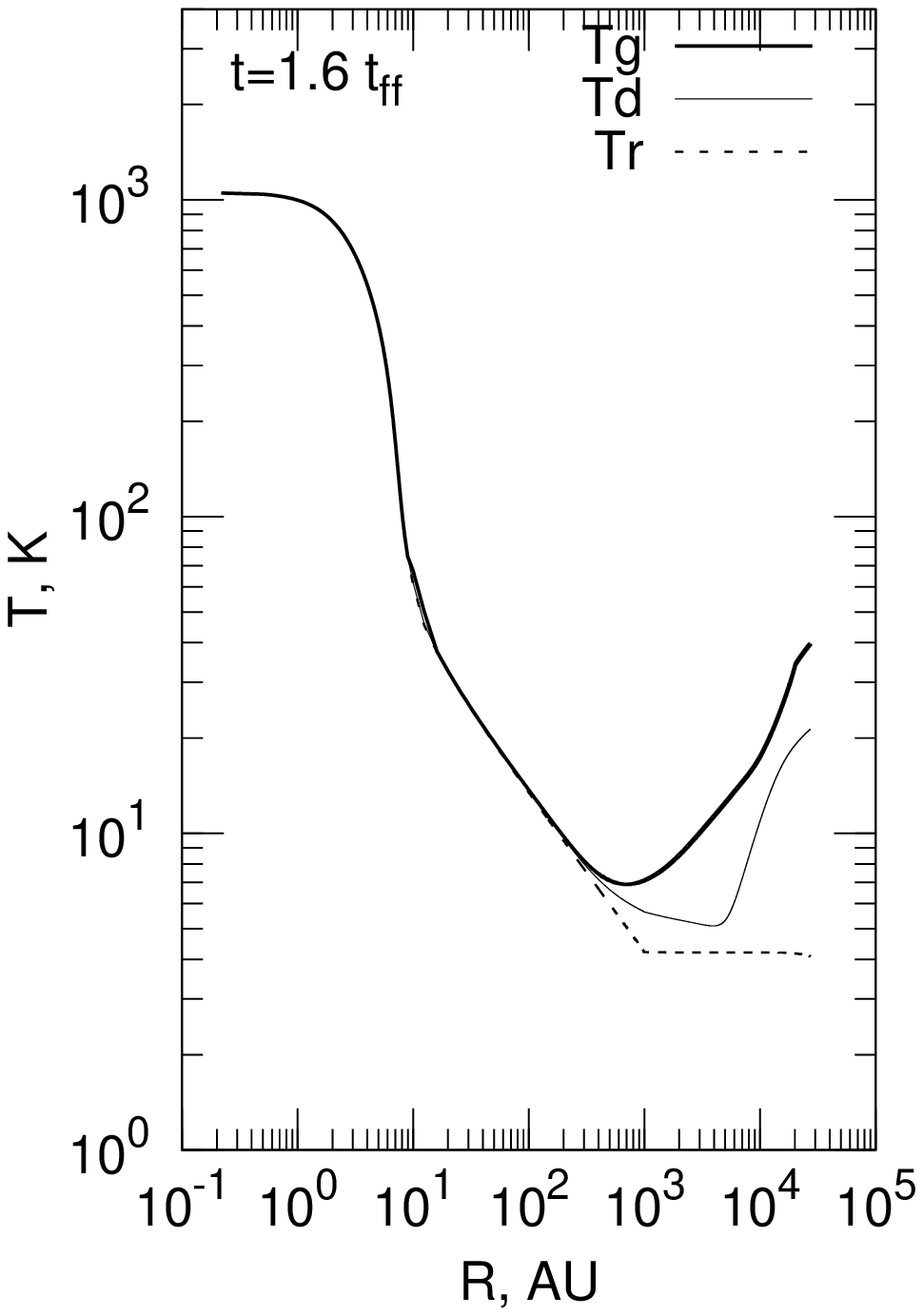}\\
    \includegraphics[scale=0.7]{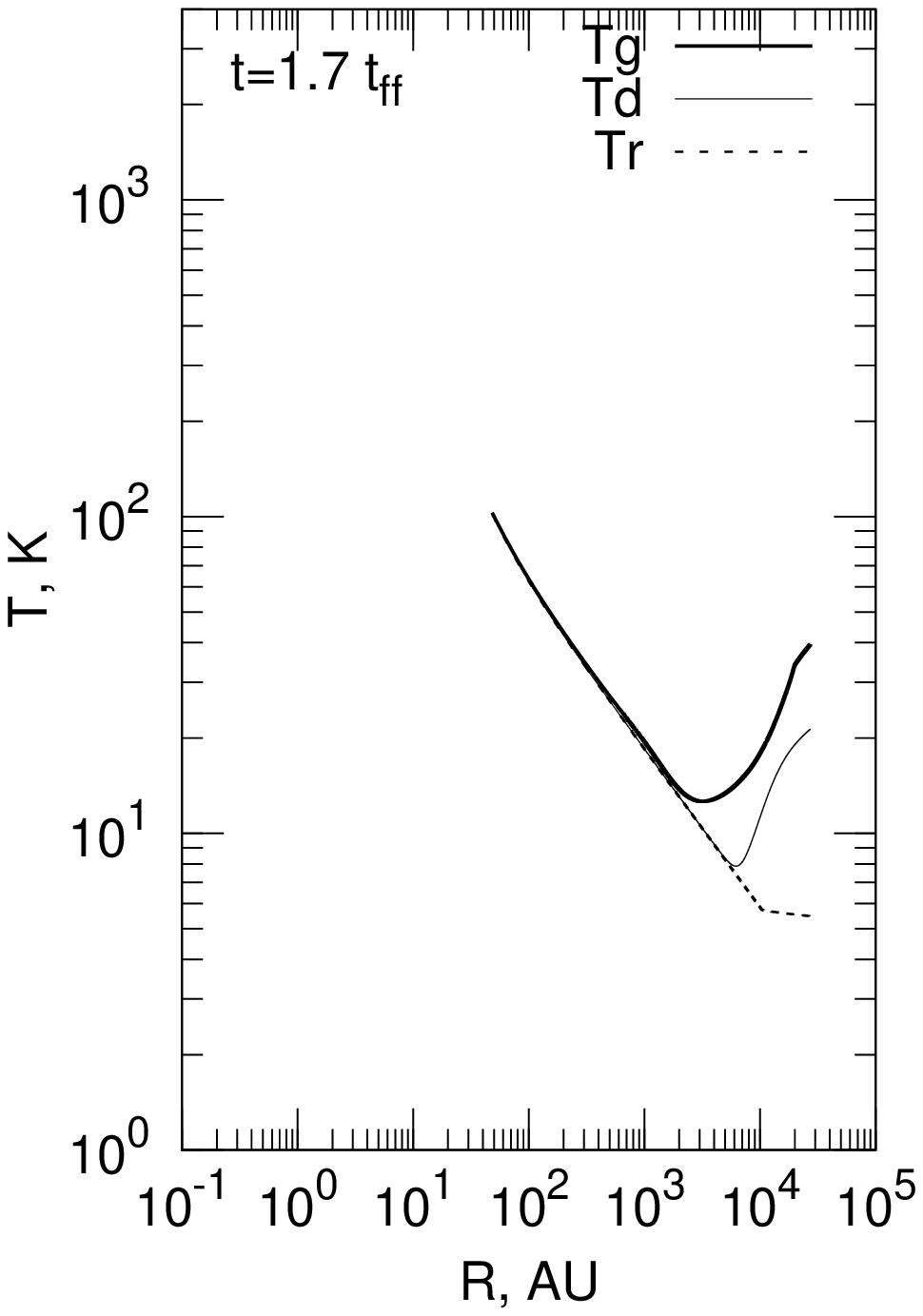}
    \includegraphics[scale=0.7]{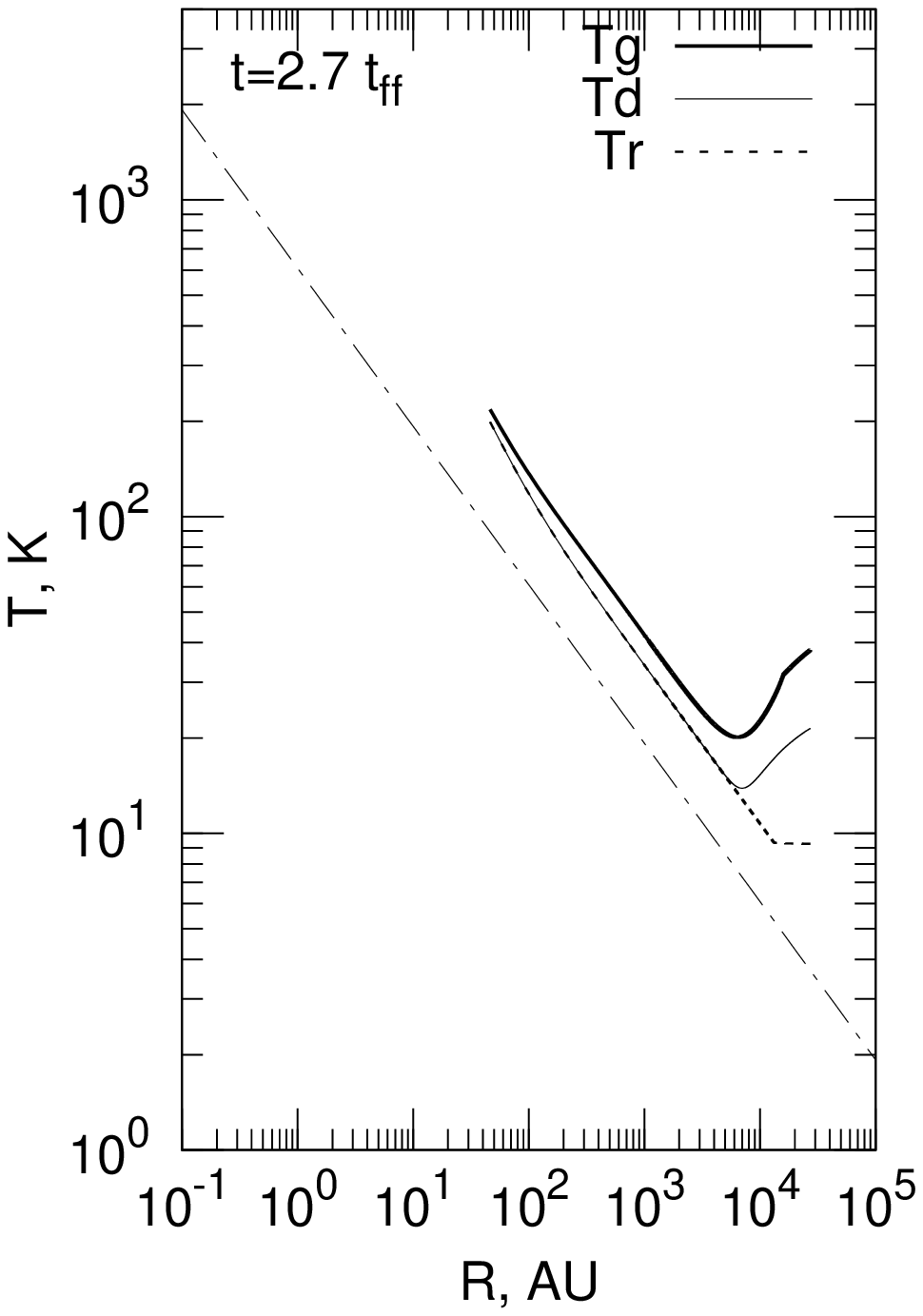}
    \caption{Distributions of the gas ($T_{\textrm{g}}$), dust
($T_{\textrm{d}}$), and radiation ($T_{\textrm{r}}$) temperatures
at various times during the evolution of a protostellar cloud. The
dash–dot line corresponds to the temperature distribution of the
stellar radiation in the absence of absorption. \hfill}
\end{figure*}

\section{EVOLUTION OF THE CLOUD WITH TRANSITION TO THE ACCRETION PHASE}

Figures~1 and~2 present the results of our modeling of the
evolution of a protostellar cloud with the initial conditions
taken from the model described in~[2]. The distributions of the
hydrogen number density $n(\textrm{H}_2)$, velocity $v$, and the
temperatures of the gas ($T_{\textrm{g}}$), dust
($T_{\textrm{d}}$, and IR radiation ($T_{\textrm{r}}$) are shown
for four times. Let us describe the main features of these
distributions for each of these times.

1. \underline{$t=0.0~t_{\textrm{ff}}$.} The initial protostellar
cloud with parameters identical to those described in~[2]. The
distributions of the density and temperature were obtained for a
cloud with central density $n(\textrm{H}_2)=10^5$~cm$^{-3}$ in
hydrostatic and thermal equilibrium. The cloud was irradiated by
external blackbody radiation with the temperature $10^4$~K and
dilution $10^{-14}$. To initiate the contraction, the density in
the hydrostatic cloud was increased by a factor of two. The
temperature of the dust and gas in the initial configuration are
different throughout volume, due to the relatively low density of
the medium. The thermal structure is governed totally by the
external radiation.

2. \underline{$t=1.6~t_{\textrm{ff}}$.} The first hydrostatic
core. As a result of the cloud compression, the central hydrogen
number density has reached $10^{14}$~cm$^{-3}$, and the
temperature 1000~K. While the temperatures of the gas and dust in
the inner regions ($r~<~300$~AU) are equal, they remain different
in the outer parts of the cloud. The temperature of the IR
radiation is close to the temperature of the gas and dust inside
$r<300$~AU, indicating that this region is opaque to its own
thermal radiation. The quasi-hydrostatic equilibrium of the inner
region can be inferred from the velocity profile: the contraction
rate is close to zero at $r<10$~AU. This epoch corresponds to the
time when the sink cell was introduced.

3. \underline{$t=1.7~t_{\textrm{ff}}$.} The time $0.1~t_{\textrm
{ff}}$ after the formation of the protostar and introduction of
the sink cell. The temperatures of the gas, dust, and radiation
are equal within $r~<~1000$~AU, and reach 100~K at the inner
boundary. At the same time, the temperatures in the envelope are
close to those in the prestellar phase of the evolution. The
hydrogen number density and gas velocity at the inner boundary are
$10^8$~cm$^{-3}$ and 2~km/s.

4. \underline{$t=2.7~t_{\textrm{ff}}$}. The protostellar phase,
corresponding to a developed accretion regime. At this time, the
mass of the star is 3~$M_{\odot}$, its age is 90000~yr, its radius
11~$R_{\odot}$, the accretion rate $3.6\times
10^{-5}$~$M_{\odot}$/yr, the photospheric luminosity
64~$L_{\odot}$, and the accretion luminosity 155~$L_{\odot}$. The
distributions of the density and velocity are monotonic and are
described well by the power laws $\rho \propto r^{-3/2}$ and $v
\propto r^{-1/2}$, whose indices are characteristic for the
self-similar solutions in the case of spherically symmetric
accretion~[7] (they are shown by the dash.dot line in Fig.~1). At
the border of the sink cell, the hydrogen number density is close
to $10^8$~cm$^{-3}$, the accretion velocity has increased to
$\approx 10$~km/s. The temperature of the gas and dust in the
inner regions is governed by the IR radiation of the protostar,
while it continues to be determined by the interstellar radiation
at the outer boundary of the cloud. The dash–dot line in Fig.~2
shows the distribution of the temperature of the stellar radiation
in the absence of absorption, calculated using the expression
$T(r) = \left(\dfrac{L_{\textrm{cell}}}{4\pi a
c}\right)^{1/4}r^{-1/2}$. The model distribution is located
slightly above this dependence, due to the finite optical depth of
the envelope and the additional heating due to adiabatic
contraction.

\section{APPROXIMATE METHOD OF COMPUTATION OF THERMAL EVOLUTION}

The mathematical model describing the thermal structure of the
protostellar cloud is fairly complex. The direct realization of
the method used to numerically solve the corresponding system of
equations in the multi-dimensional case involves the solution of a
number of additional problems of a technical nature. In
particular, a highly sparse $M\times M$ matrix must be inverted,
where $M$ is the total number of cells in the computational grid.
The classical methods for solving such problems (the method of
Gauss, LU decomposition, etc.) are very inefficient. Complex
iterative algorithms are used, such as multi-grid methods, the
methods of Krylov, etc.~[24]. The final choice of the method used
in the multi-dimensional case should be determined by the desired
compromise between accuracy, computational time, and complexity of
the realization of the method. The use of adequate approximate
methods in the multi-dimensional calculations is an attractive
idea. We devised an approximate method for the solution of the
system of equations describing the thermal structure of the
protostellar cloud, which we call the ``adaptive
$\tau$-approximation''. In this method, the diffusion operator in
the equation for the IR energy is replaced by the local
(coordinate dependent) algebraic operator

We devised an approximate method for the solution of the system of
equations describing the thermal structure of the protostellar
cloud, which we call the ``adaptive $\tau$-approximation''. In
this method, the diffusion operator in the equation for the IR
energy is replaced by the local (coordinate dependent) algebraic
operator

\begin{equation}\label{eq-tau1}
\nabla \cdot \left( \frac{1}{\sigma_{\textrm{R}}} \nabla
E_{\textrm{r}} \right) = {-}\frac{E_{\textrm{r}} -
E_{\textrm{bg}}}{\tau}.
\end{equation}

Here, $E_{\textrm{bg}}$ is the energy density of the background IR
radiation and $\tau$ the characteristic diffusion time of the IR
radiation. In essence, the adaptive $\tau$-approximation describes
the release of IR radiation energy at a given point over a
characteristic local timescale $\tau$, which is determined by the
distribution of the radiation sources and the diffusion
coefficient in the cloud. Generally speaking, various procedures
can be used to calculate $\tau$ in the algebraic operator. We
chose $\tau$ so that the stationary solution obtained using the
$\tau$-approximation coincides with the exact solution.

The coefficient $\tau$ is computed as follows. Consider a
stationary distribution of the gas and dust temperatures and the
IR and UV energy. In this case, the subsystem of equations
describing the thermal structure of the cloud can be reduced to
the equation

\begin{equation}\label{eq-tau2}
\nabla \cdot \left( \frac{1}{\sigma_{\textrm{R}}} \nabla
\bar{E}_{\textrm{r}} \right) = Q,
\end{equation}
where
\begin{equation}\label{eq-tau3}
Q = \Lambda_{\textrm{g}} - \Gamma_{\textrm{g}} - S.
\end{equation}

Here, $\Lambda_{\textrm{g}}$ is the gas cooling function via the
emission of molecular line radiation, $\Gamma_{\textrm {g}}$ the
gas heating function via cosmic rays and photoelectrons, and $S$
the dust heating function via external UV radiation. A bar denotes
the steady-state distribution of the energy density. Solving this
equation taking into account of the necessary boundary conditions
yields an approximate distribution for the energy density of the
IR radiation. The coefficient $\tau$ can then be found using~(20)
and the corresponding approximate equation

\begin{equation}\label{eq-tau4}
-\frac{\bar{E}_{\textrm{r}} - E_{\textrm{bg}}}{\tau} = Q.
\end{equation}

Technically, the adaptive $\tau$-approximation method can be
implemented substantially more easily than the “full” method
described above. Replacing the diffusion operator with an
algebraic operator leads to a local nonlinear system of equations
for determining $T_{\textrm{g}}^{n+1}$, $T_{\textrm{d}}^{n+1}$,
and $T_{\textrm{r}}^{n+1}$. In turn, the finite-difference
equation (21) is linear ($\sigma_{\textrm{R}}$ and $Q$ depend on
the previous time step) and its solution is appreciably simpler
than the solution of the general system of equations.

Consider the spherically symmetric case, assuming that the energy
density of the IR radiation is equal to the background energy
density $E_{\textrm{bg}}$ at the outer boundary of the cloud, at
$r = R$. At the inner boundary of the cloud, $r = r_{1}$, the flux
of IR radiation from the central protostar is specified to be $F =
F_{\textrm{cell}}$. Integration of equations~(21) and~(23) yields

\begin{equation}\label{eq-tau5}
 \tau = \frac{1}{Q} \left\{
 \int\limits_{r}^{R} d\xi \frac{\sigma_{\rm R}(\xi)}{\xi^2}
 \left[
 \int\limits_{r_{1}}^{\xi} d\eta \eta^2 Q(\eta) +
 r_{1}^2 F_\text{cell} \right]
 \right\}.
\end{equation}

Note that the IR luminosity of the protostar is then
$L_{\textrm{cell}} = 4\pi {r_{1}}^2 F_{\textrm{cell}}$. In the
case when $\sigma_{\textrm{R}} = {\textrm{const}}$ and $Q =
{\textrm{const}}$, we have

\begin{equation}\label{eq-tau6}
 \tau = \frac{r_{1}^2 F_\text{cell}}{Q} \sigma_{\rm R}
 \left( \frac{1}{r} - \frac{1}{R} \right) +
 \frac{1}{3} \sigma_{\rm R} \left[
 \frac{R^2 - r^2}{2} - r_{1}^3 \left( \frac{1}{r} - \frac{1}{R} \right)
 \right].
\end{equation}

For regions sufficiently far from the cloud boundary, we can
assume $r_{1} \ll R$ and $r \ll R$. Therefore, we can obtain
from~(25) the approximate expression

\begin{equation}\label{eq-tau7}
\tau \approx \frac{r_{1}^2 F_{\textrm{cell}}}{Q}
\frac{\sigma_{\textrm{R}}}{r} + \frac{1}{6} \sigma_{\textrm{R}}
R^2.
\end{equation}

Thus, the possibility of taking into account the additional source
of IR radiation due to the central protostar is an important
advantage of the adaptive $\tau$-approximation method. This makes
it possible to use this method to calculate the thermal structure
of a protostellar cloud in the multi-dimensional case with the
constructed model for the protostar.

The adaptive $\tau$-approximation method gives the exact solution
(or close to it) in two important special cases. In the initial
stages of the contraction of the protostellar cloud, the medium is
optically thin to IR radiation. This means that the timescale
$\tau$  is much shorter than the dynamical timescale, which can be
estimated as the cloud free-fall timescale $t_{ff}$. As a result,
the diffusion operator dominates, and we obtain nearly the
steady-state energy density of the IR radiation, close to the
background value $E_{\textrm {bg}}$. Later, $\tau$ increases
proportional to the density in the central regions of the cloud,
while dynamical timescale decreases with increasing density
proportional to $\rho^{-1/2}$. Therefore, in late stages of the
contraction, the situation will become exactly the opposite:
$\tau$ will be much longer than the dynamical timescale. In this
case, the diffusion operator is not important, and the energy
density of the IR radiation will vary mainly due to the source
terms. Therefore, we expect that the $\tau$-approximation will
yield a solution close to the solution obtained using the rigorous
diffusion operator.

\begin{figure*}[p!]
    \begin{center}
    \includegraphics[scale=0.7]{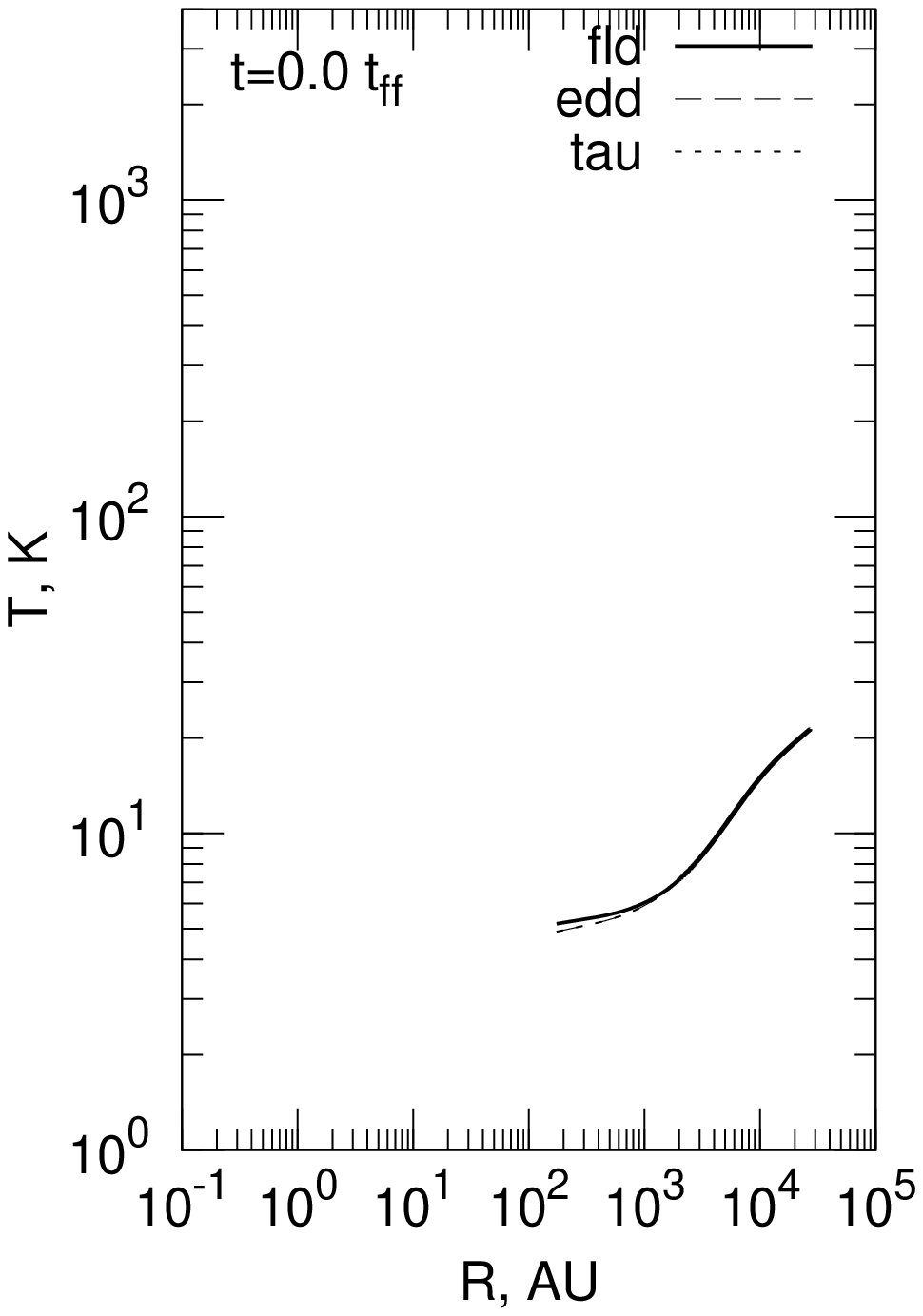}
    \includegraphics[scale=0.7]{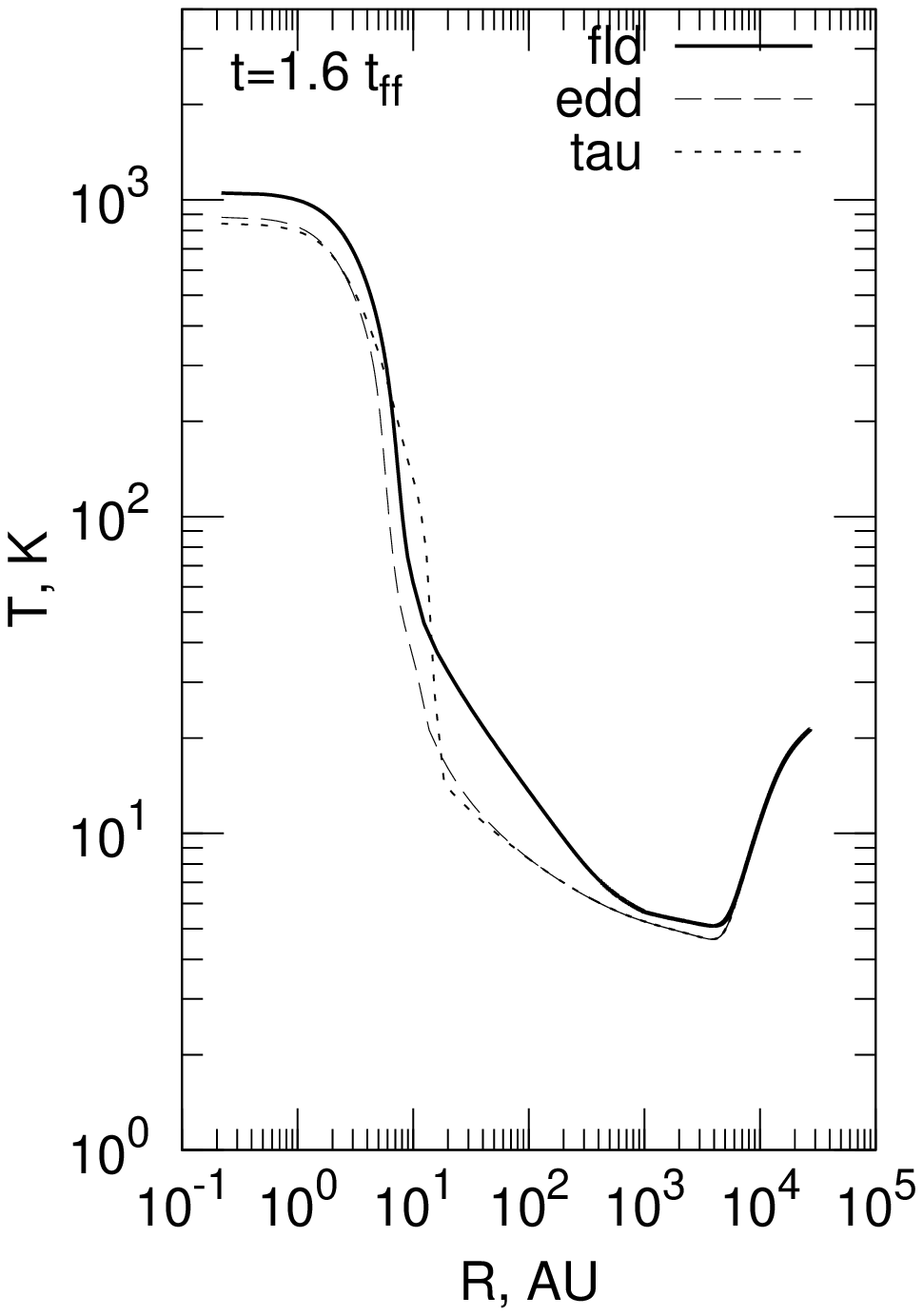}\\
    \includegraphics[scale=0.7]{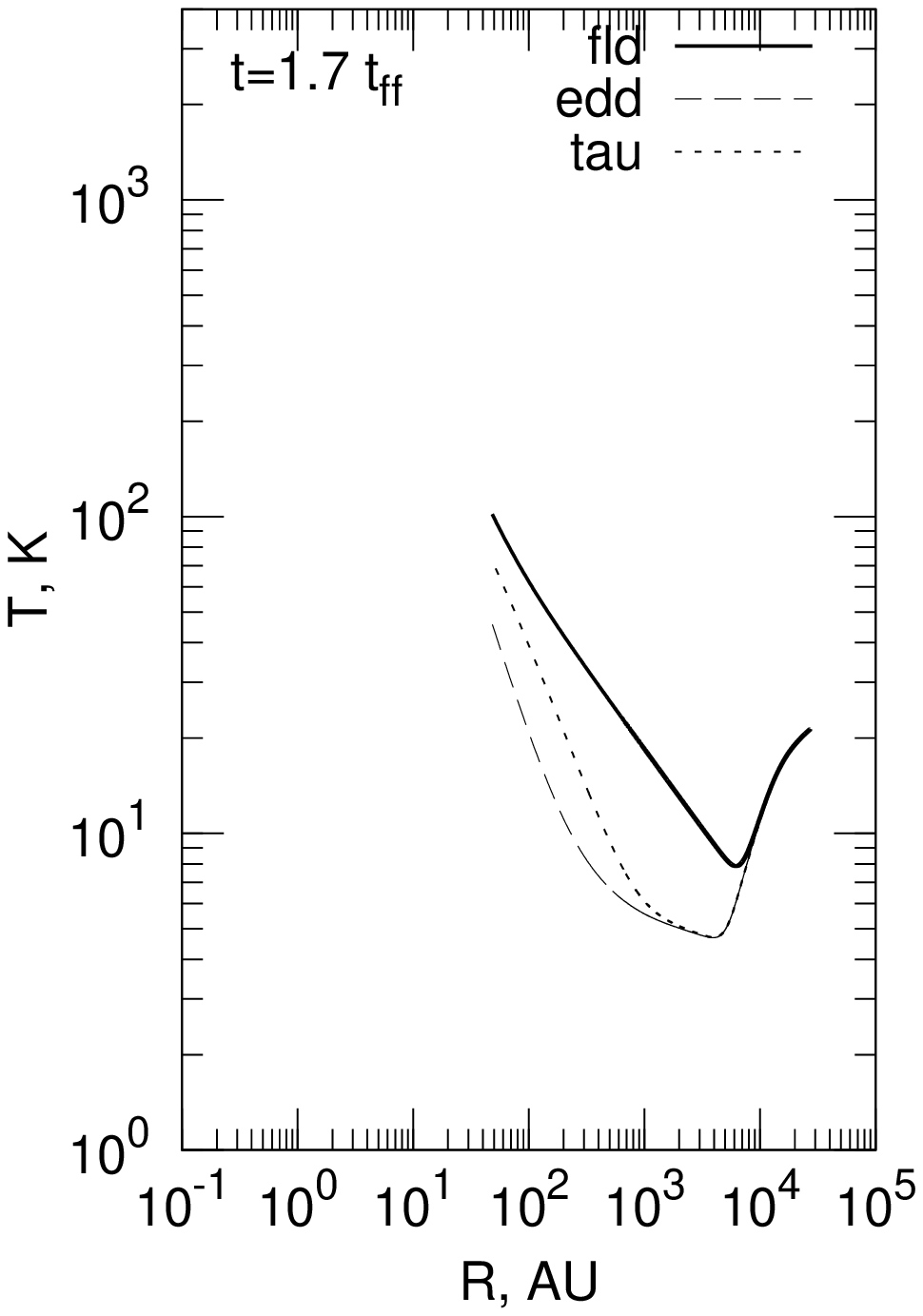}
    \includegraphics[scale=0.7]{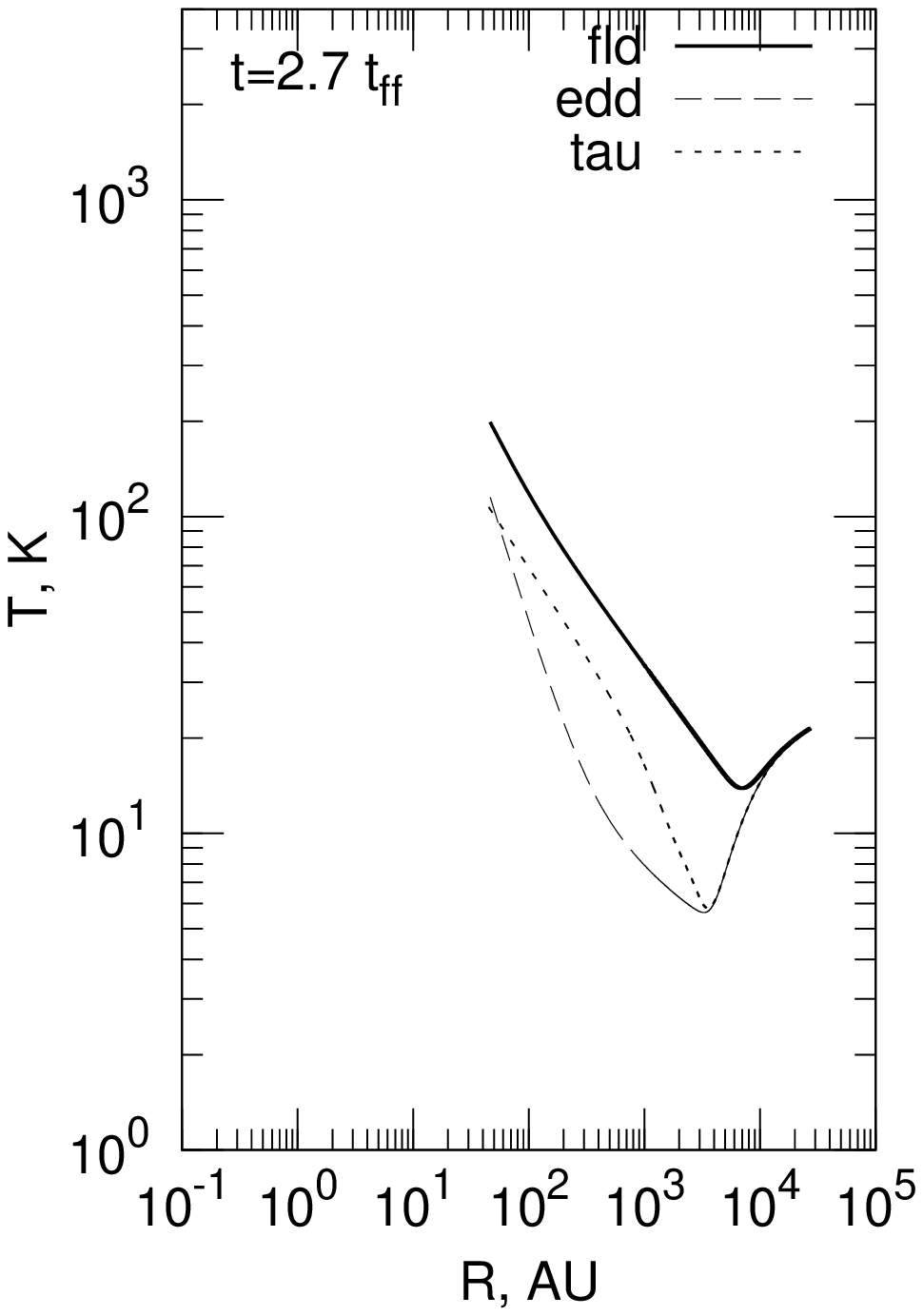}
    \end{center}
    \caption{Distributions of the dust temperature $T_{\textrm{d}}$
    for various epochs during the evolution of a protostellar cloud,
    obtained using the $\tau$-approximation (tau),
    Eddington approximation (edd), and the diffusion approximation
    with a flux limiter (fld).
\hfill}
\end{figure*}

Figure 3 shows the distribution of the dust temperature $
T_{\textrm{d}}$ for various epochs in the evolution of a
protostellar cloud obtained using the $\tau$-approximation (tau),
Eddington approximation (edd), and diffusion approximation with a
flux limiter (fld) (for a description of the Eddington method and
the diffusion approximation with a flux limiter, see Section~2).
The $\tau$-approximation and Eddington approximation agree fairly
well with the ``fld'' method in the prestellar phase, up to the
time of the formation of the hydrostatic core. Significant
differences between the distributions are observed in the
accretion phase. The decrease in the dust temperature at
100--1000~AU is stronger in the Eddington approximation than in
the ``FLD'' approach. This is associated with the ill-posed nature
of the Eddington approximation for the streaming regime. The
temperature obtained is higher in the $\tau$-approximation than in
the Eddington approximation, but lower than in the ``FLD''
approximation. Although the adaptive $\tau$-approximation yields a
significant error, it reproduces the behavior of the temperature
fairly well, and is suitable for simple (preliminary) calculations
of the evolution of protostellar clouds, as well as calculations
of the evolution of optically thick protostellar disks.

\section{COMPARISON WITH OBSERVATIONS OF PRESTELLAR AND PROTOSTELLAR CORES}

The main aim of the thermal model presented here is a direct
comparison of the results of the dynamical calculations with
observations. The Herschel Space Observatory has obtained rich
observational data on various protostellar objects in recent
years. In particular, the spatial and thermal structure of a
number of dense molecular clouds, including both prestellar and
protostellar cores, were studied in [25–-27]. The distributions of
the surface number density of hydrogen
$N_{\textrm{H}}^{\textrm{obs}}$ and the dust temperature averaged
over the line of sight $\tilde{T}^{\textrm{obs}}_{\textrm{d}}$ for
a number of prestellar and protostellar cores are presented
in~[27]. It is natural for us is to attempt to reproduce these
distributions using our model. Two circumstances must be taken
into account when modeling the distributions of $N_{\textrm{H}}$
and $\tilde{T}_{\textrm{d}}$ based on the theoretical
distributions $n_{\textrm{H}}$ and $T_{\textrm{d}}$. First, the
temperature is not an additive quantity, and the averaging along
the line of sight must be carried out consistently with the method
used in~[27] to reconstruct the distributions. Second,
$N_{\textrm{H}}^{\textrm{obs}}$ and
$\tilde{T}^{\textrm{obs}}_{\textrm{d}}$ were determined from
observations of distributions of the radiation intensity with
finite angular resolution, $\textrm{HPBW}=36^{\prime\prime}$.

Let us consider the radiation intensity $I_{\nu}$ in the optically
thin approximation:

\begin{equation}
I_{\nu} = m_{\textrm{H}}\kappa_{\nu}
\int\limits_{0}^{N_{\textrm{H}}} B_{\nu}(T_{\textrm{d}})  dN,
\label{eq_Inu}
\end{equation}

where $\kappa_{\nu}$ [cm $^2$/g] is the opacity per gram of gas,
$m_{\textrm{H}}$ the atomic mass of hydrogen, $N$ the surface
number density of hydrogen along the line of sight,
$N_{\textrm{H}}$ the total surface number density of hydrogen, and
$B_{\nu}$ the Planck function. In essence, it was assumed in~[27]
that

\begin{equation}
I_{\nu}^{\textrm{obs}} =m_{\textrm{H}}\kappa_{\nu}
B_{\nu}(\tilde{T}_{\textrm{d}}^{\textrm{obs}})
N_{\textrm{H}}^{\textrm{obs}}. \label{ralf}
\end{equation}

In other words, distributions
$\tilde{T}_{\textrm{d}}^{\textrm{obs}}$ and
$N_{\textrm{H}}^{\textrm{obs}}$ are found for which the spatial
distribution (28) satisfactorily describes the observed maps at
all available frequencies. Since $I_{\nu}^{\textrm{obs}}$is a
convolution of the true observed radiation intensity with the
telescope beam, the distribution.
$\tilde{T}^{\textrm{obs}}_{\textrm {d}}$ d is a result of
averaging not only along the line of sight, but also along the
angular coordinate.

To construct the theoretical distribution
$\tilde{T}_{\textrm{d}}^{\textrm{conv}}$, we computed the
distribution of the radiation intensity $I_{\nu}(x,y)$ using
Eq.~(27) and convolved it with the Gaussian telescope beam
$W(x,y)$:

\begin{equation}
I_{\nu}^{\textrm{conv}}(x,y) = {}\\
\nonumber{}= \int W(x-x^{\prime},y-y^{\prime})
I_{\nu}(x^{\prime},y^{\prime}) dx^{\prime} dy^{\prime}.
\end{equation}

In this formula, $W(x,y) = \dfrac{1}{\pi H^2} \exp
\left(-\dfrac{x^2+y^2}{H^2}\right)$, where
$H=\dfrac{\textrm{HPBW}}{2\sqrt{\ln 2}}$, and HPBW is the
half-power beam width in radians. We compared
$\tilde{T}_{\textrm{d}}^{\textrm{obs}}$ with the temperature .
$\tilde{T}_{\textrm{d}}^{\textrm{conv}}$ found from the expression

\begin{equation}
B_{\nu}(\tilde{T}_{\textrm{d}}^{\textrm{conv}}) =
\dfrac{I_{\nu}^{\textrm{conv}}}{m_{\textrm{H}}\kappa_{\nu}N_{\textrm{H}}^{\textrm{conv}}}
. \label{eq_Tmean}
\end{equation}

Averaged over the telescope beam, the surface number density is

\begin{equation}
N_{\textrm{H}}^{\textrm{conv}} = {}\\
\nonumber{}= \int W(x-x^{\prime},y-y^{\prime})
N_{\textrm{H}}(x^{\prime},y^{\prime}) dx^{\prime} dy^{\prime} .
\end{equation}

Note that the method for averaging the temperature described above
does not require knowledge of the opacity $\kappa_{\nu}$ in~(27)
and~(30), since it cancels out in the final formula. However, the
method does depend on the wavelength used in the Planck function.
We used a wavelength of 160~mkm, close to the maximum of the
observed radiation flux. We compared the model with the observed
distributions averaged over the sample of sources shown in Fig. 8
of [27] by blue and red lines. When calculating the distributions
$N_{\textrm{H}}^{\textrm{conv}}$ and $\tilde{T}^{\textrm{conv}}$,
we assumed that the cloud was located at a distance of 140~pc, and
that the telescope beam was Gaussian with
$\textrm{HPBW}=36^{\prime\prime}$.

\begin{figure*}[t!]
    \begin{center}
    \includegraphics[scale=0.6]{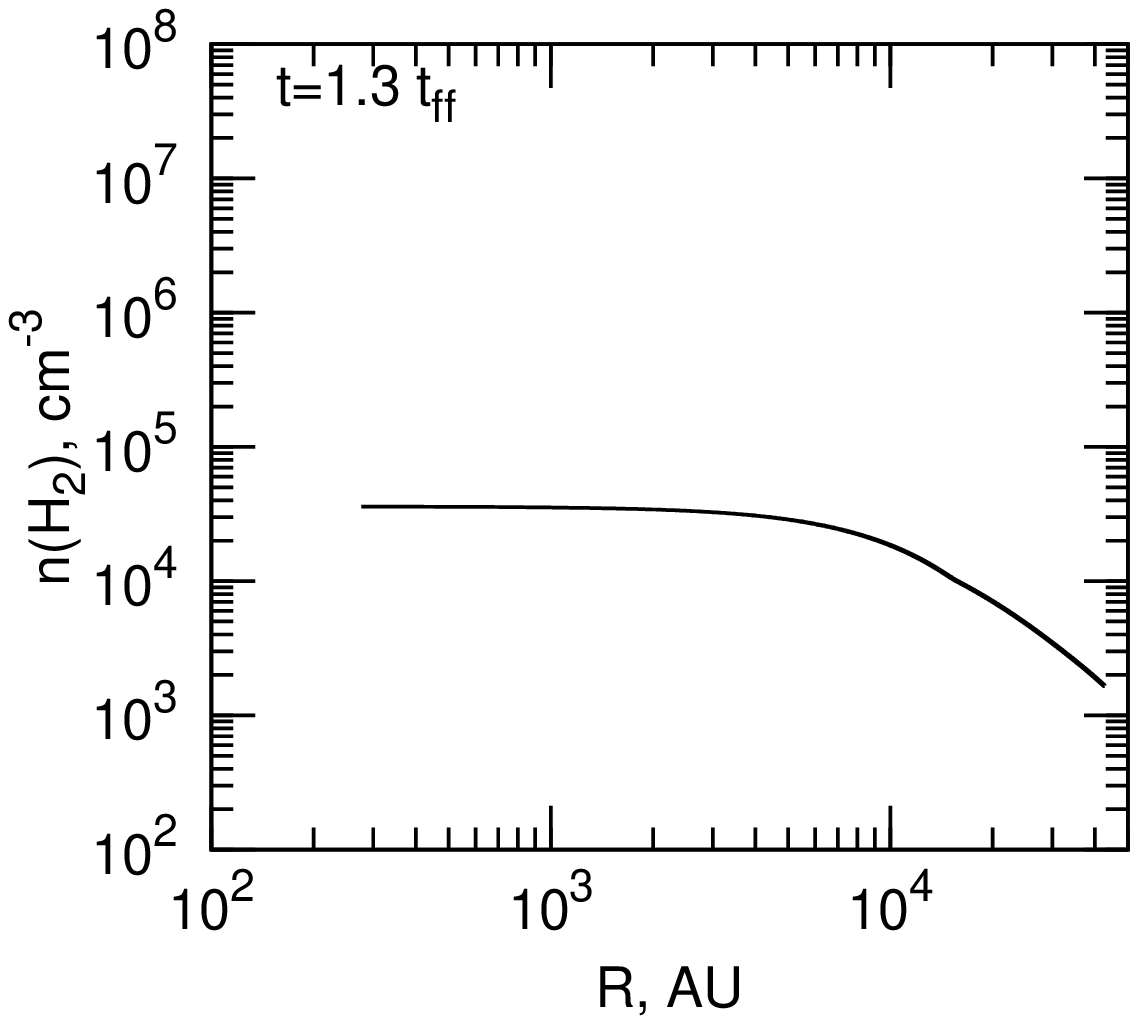}
    \includegraphics[scale=0.6]{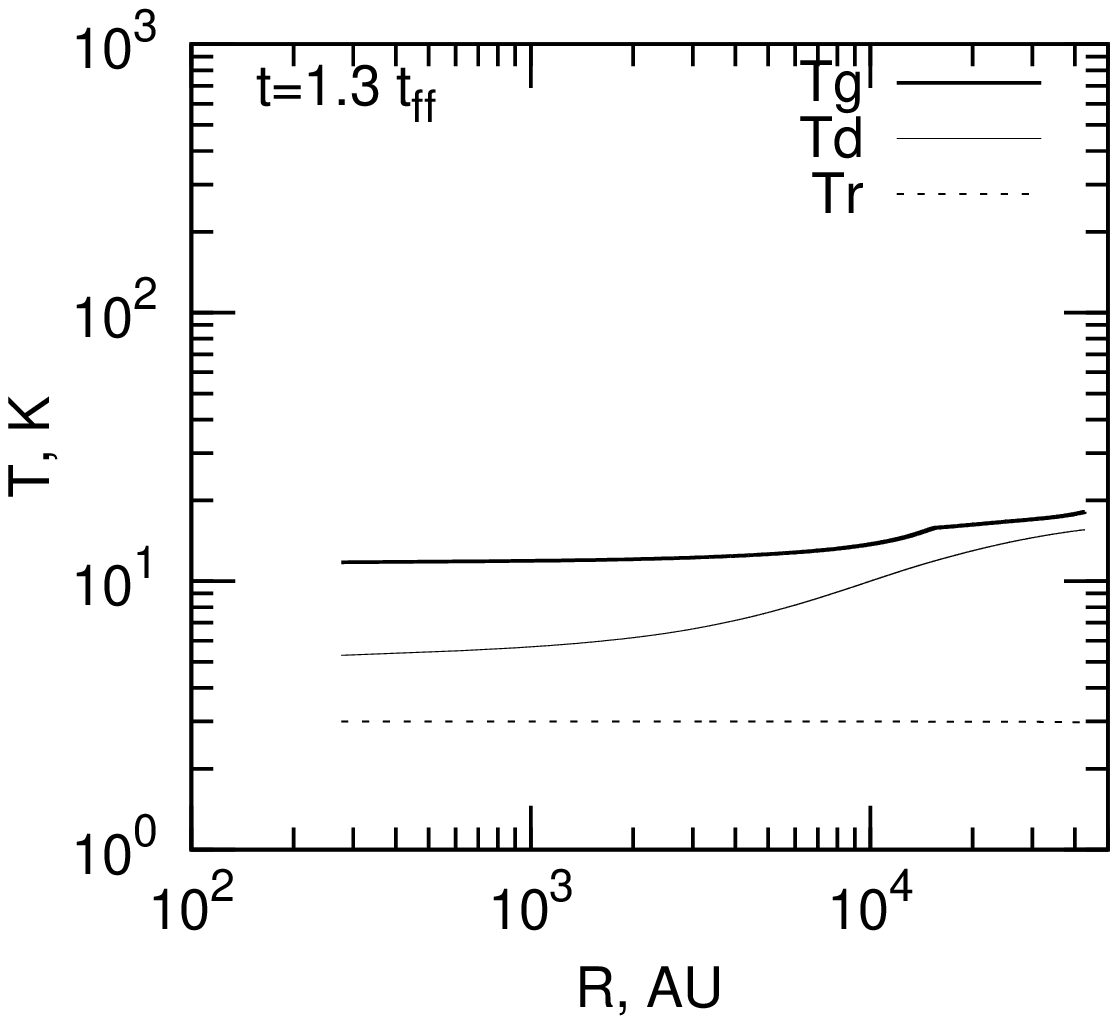} \\
    \includegraphics[scale=0.6]{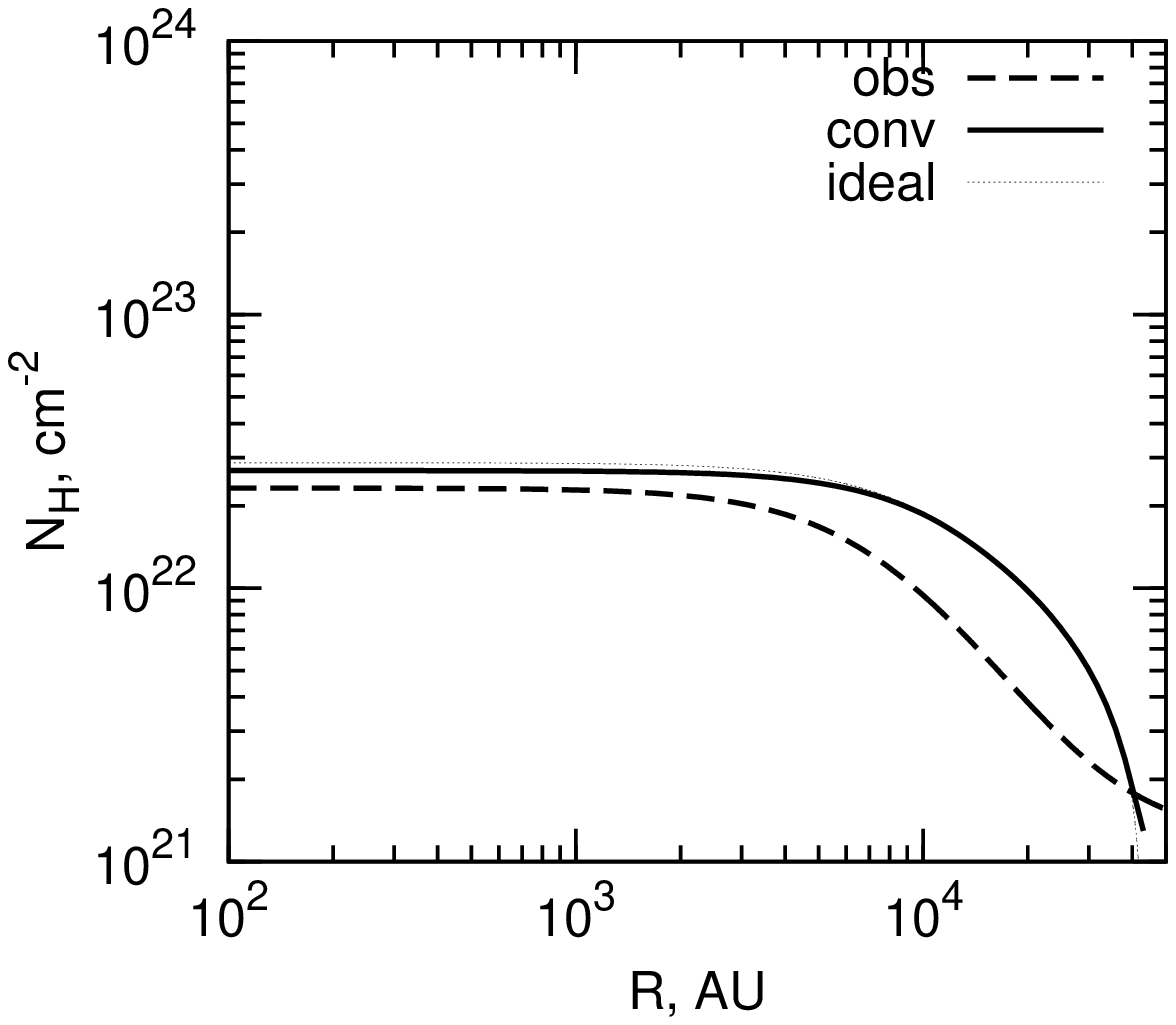}
    \includegraphics[scale=0.6]{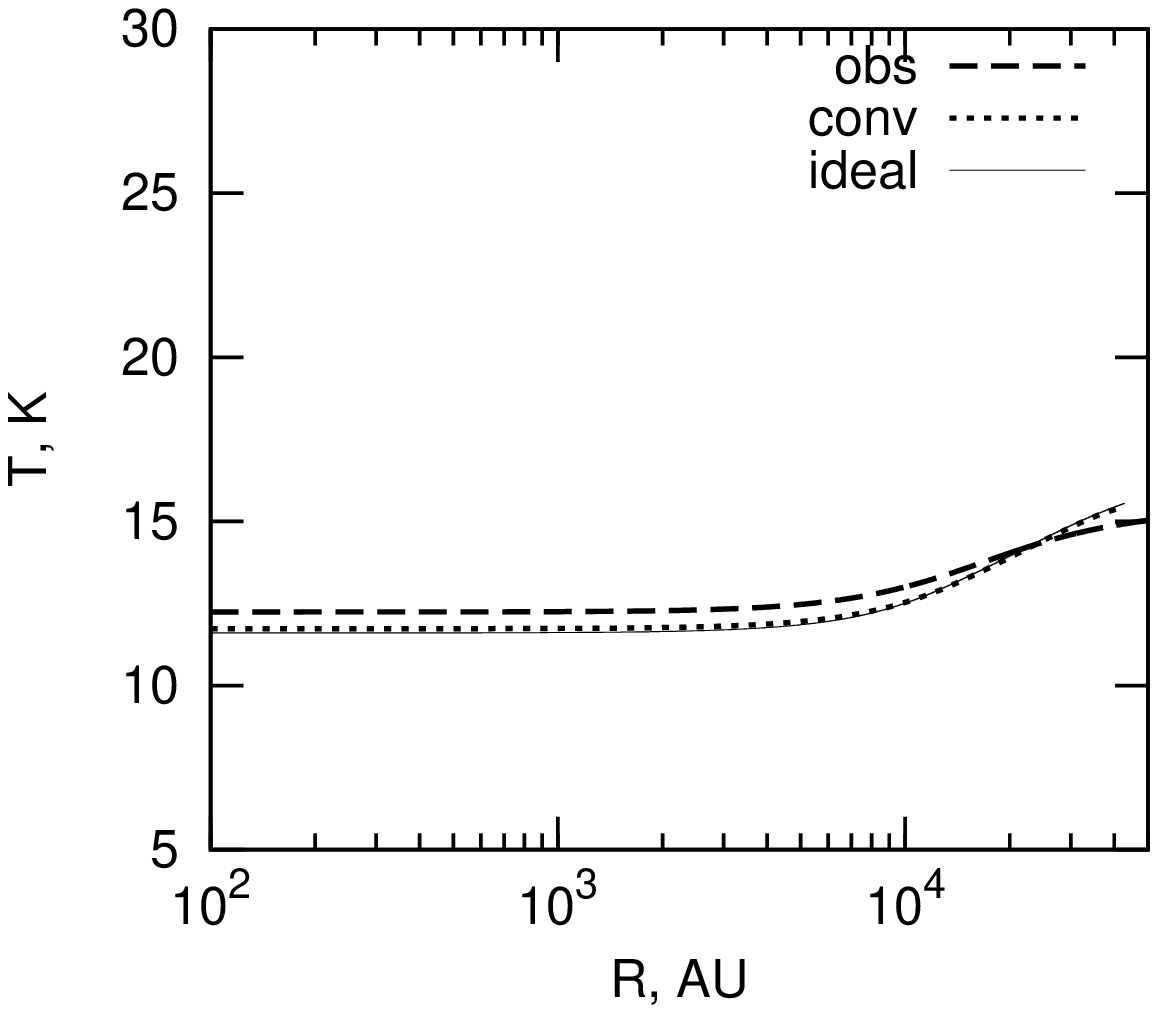}
\end{center}
    \caption{Distributions for the prestellar core:
    hydrogen number density
(upper left); gas, dust, and radiation temperature (upper right);
surface density (lower left); and dust temperature weighted over
the line of sight (lower right). In the lower panels, the model
distributions are shown by the solid curves and the observed
distributions from~[27] by the bold dashed curves. The model
distributions were derived by convolving the ideal distributions
(shown by the thin dashed curves) with a Gaussian beam with
$\textrm{HPBW} =36^{\prime\prime}$.
        \hfill}
\end{figure*}

\begin{figure*}[t!]
    \begin{center}
    \includegraphics[scale=0.6]{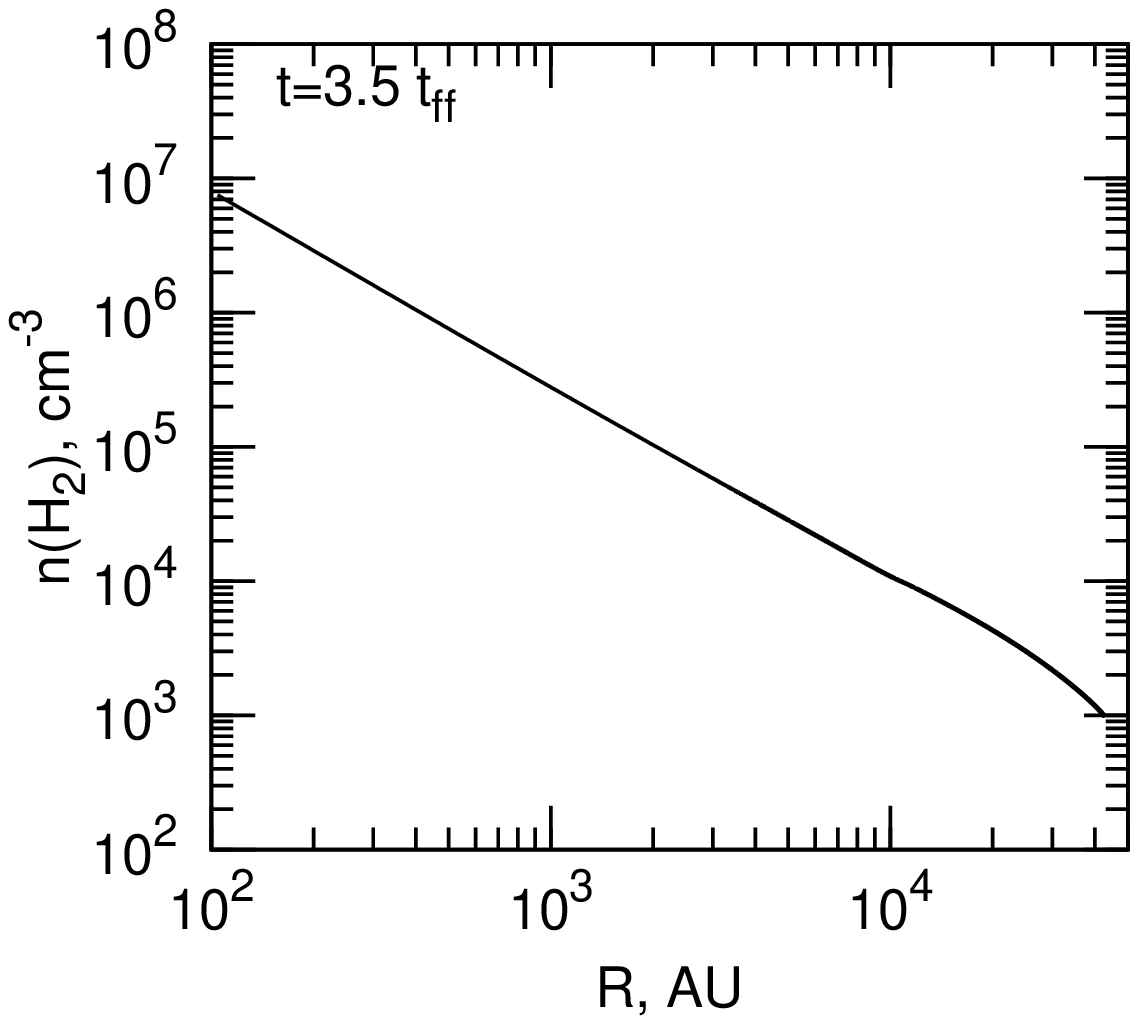}
    \includegraphics[scale=0.6]{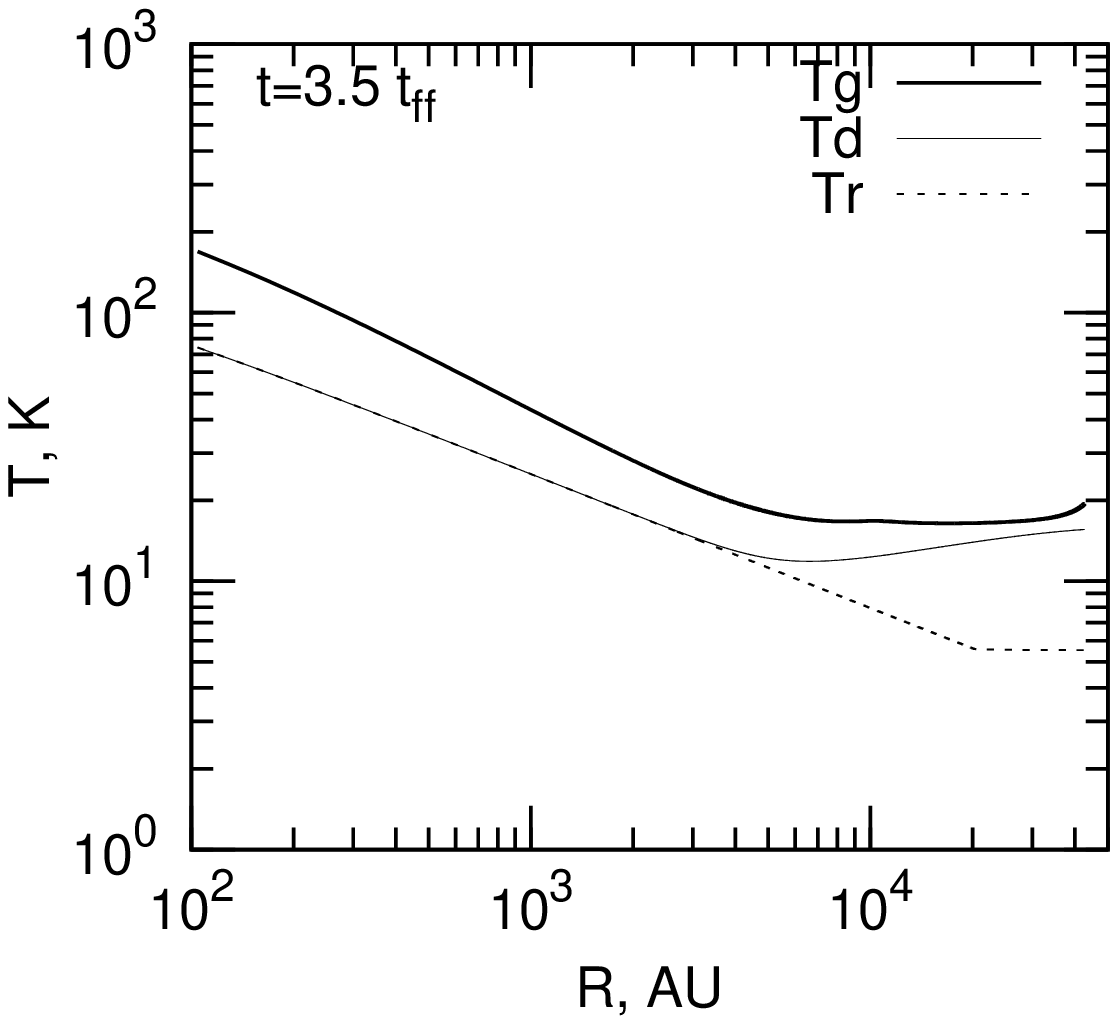} \\
    \includegraphics[scale=0.6]{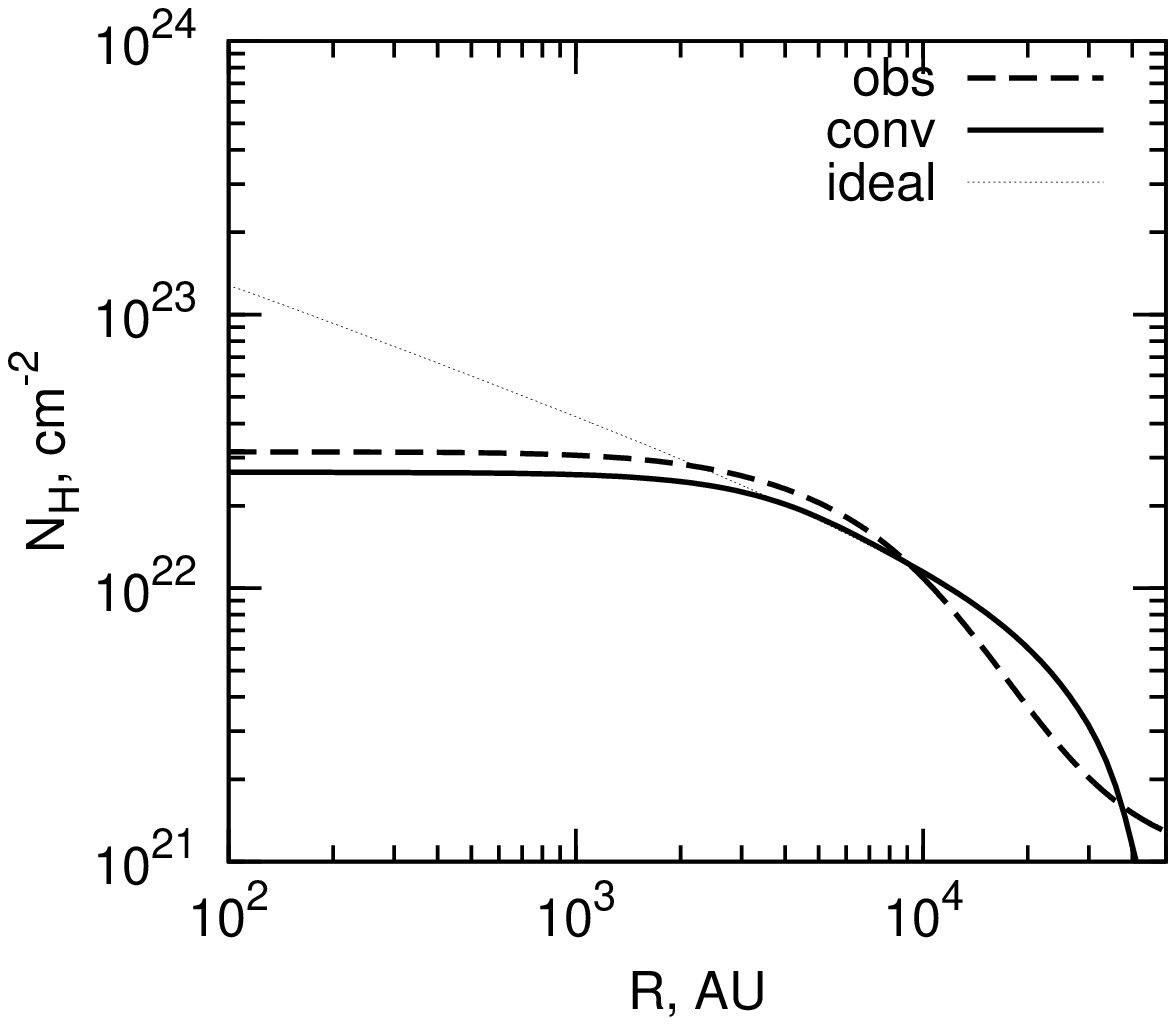}
    \includegraphics[scale=0.6]{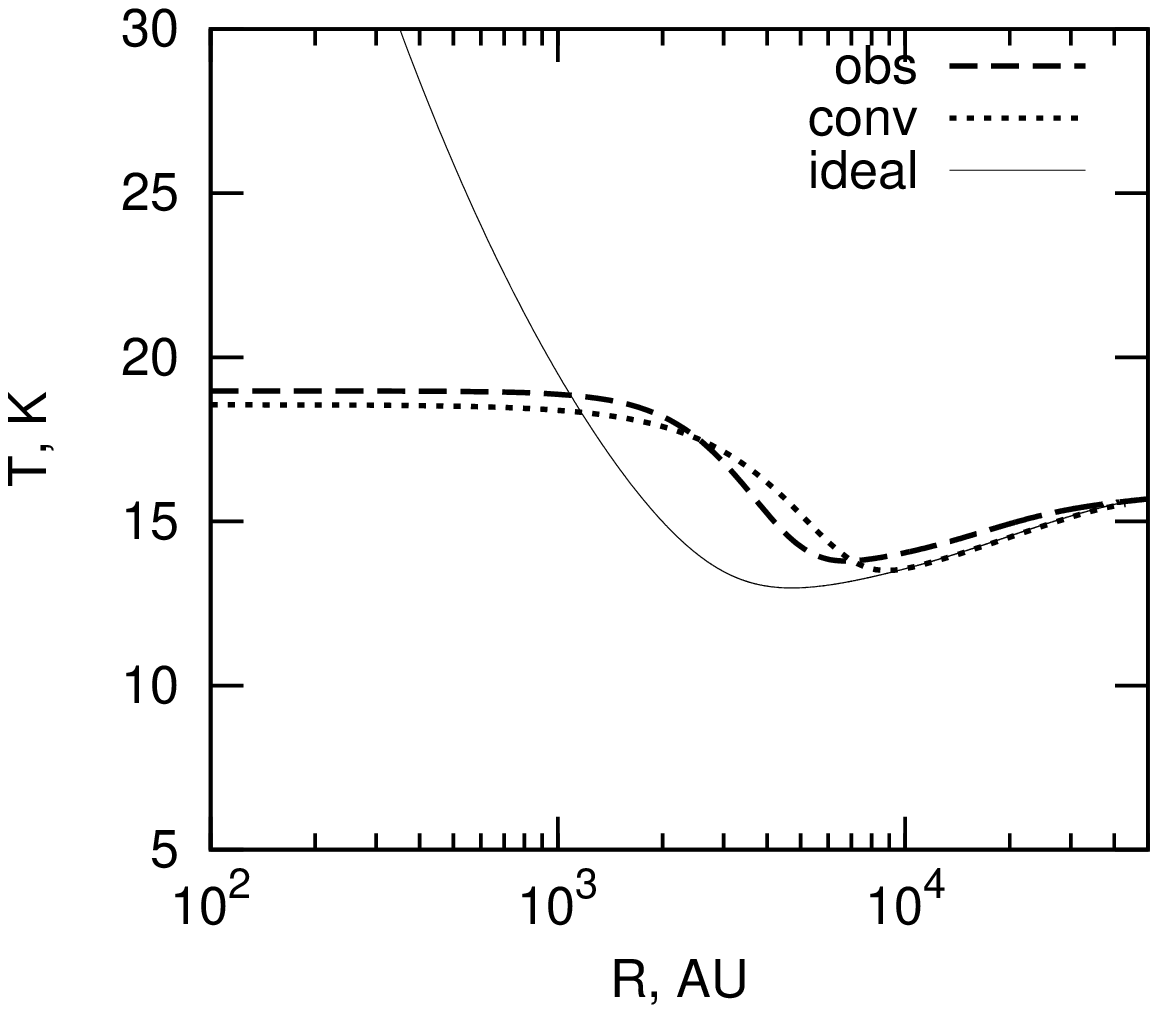}
    \end{center}
    \caption{Same as in Fig.~4 for protostellar core. \hfill}
\end{figure*}

\begin{table}[t!]
    \begin{center}
    \caption{Parameters of the prestellar cloud model}
        \begin{tabular}{l|c}
            \hline
            \multicolumn{1}{c|}{Parameter}  &  Value \\
            \hline
            Central concentration  H$_2$    &\;\;\,$4\times 10^4$~cm$^{-3}$ \\
            Radius                             &$5\times 10^4$~AU  \\
            Mass                             &\;\;\;\,6 $M_\odot$  \\
            Temperature of background UV radiation  & $10^4$~K \\
            Dilution of background UV radiation &$1.5\times 10^{-15}$\;\;\;\;\, \\
            Evolutionary time span                    &330~ka \\
            \hline
        \end{tabular}
            \end{center}
\end{table}

We suppose that the mean observed distributions of the density and
temperature for the prestellar cores correspond to \textit{cores
close to hydrostatic and thermal equilibrium}. Thus, the main
parameters of the model are the initial central density, the
radius of the cloud, the parameters of the external radiation
field, and the evolution time. We selected the model parameters
for the prestellar cores so that the value
$N_{\textrm{H}}^{\textrm{conv}}=2\times 10^{22}$~cm$^{-2}$ in the
direction toward the cloud center coincided with the observed
value, and the cloud radius was equal to the maximum value in the
distribution presented in~[27], namely, $R=5\times 10^4$~AU. The
selected parameters of the external radiation field were such that
$\tilde{T}_{\textrm{d}}^{\textrm{conv}}=15~K$ at the  cloud
boundary. The evolution time for the model was $1.3~t_{\textrm
{tff}}$; during this time, the central density doubled compared to
the initial state. Note that the model parameters were not
rigorously varied, and therefore are not strictly optimal, as we
did not aim to accurately reproduce the observational data. Our
goal was to determine whether our model was consistent with the
available observational data.

Figure~4 shows the distributions of the density and temperature in
the model for the prestellar core, together with a comparison
between the derived distributions $N_{\textrm{H}}^{\textrm{conv}}$
and $\tilde{T}_{\textrm{d}}^{\textrm{conv}}$ and the observed
distributions. The model parameters are given in the table.

The results were obtained in the diffusion approximation with a
flux limiter. The model and observed temperature distributions
agree both qualitatively and quantitatively. The temperature is
lower in the direction toward the center than in the direction
toward the outer region of the cloud, as a consequence of external
heating. At the same time, the surface density distributions
differ significantly. The theoretical distribution has a more
extended central plateau and steepe density decline toward the
edge. At a distance of $2\times 10^4$~AU, the model density is a
factor of three higher than the observed value. One possible
reason for this discrepancy in the observed and theoretical
surface-density distributions is that the assumption that the
prestellar cloud was initially in a quasi-equilibrium state was
too crude. Indeed, according to modern gravo-turbulent concept of
star formation~[28], prestellar cores form in the turbulent
interstellar medium. As a result of collisions of turbulent flows,
gravitationally bound fragments form, whose density distributions
may differ significantly from equilibrium distributions.

We suppose that the mean observed density and temperature
distributions for the protostellar cores correspond to
\textit{clouds that are a product of the evolution of the
prestellar cores and are in a phase of developed accretion}.
Therefore, the main parameter of the model is the evolution time.
We choose this time to be $t=3.5~t_{\textrm{ff}}$, when the
observed central surface number density
$N_{\textrm{H}}^{\textrm{conv}}=3\times 10^{22}$~cm$^{-2}$ is
reproduced. Figure~5 shows model distributions of the density and
temperature at this time together with the corresponding
distributions of $N_{\textrm{H}}$ and $\tilde{T}$ and the observed
profiles.

At this time, the age of the star in the model is 340 000 yrs, its
mass is 2.9~$~M_{\odot}$, the accretion rate is $6\times
10^{-6}$~$M_{\odot}$/yr, the photospheric luminosity of the star
is $23~L_{\odot}$, and the accretion luminosity is $45~L_{\odot}$.
Overall, the theoretical and observed temperature distributions
agree well. Note also that it is necessary to take the convolution
into account when interpreting the observations, since the model
temperature distribution unweighted by the telescope beam (thin
solid curve) differs strongly from the convolved distribution. At
the same time, small differences in the distributions of the
surface number density are observed in the outer parts of the
cloud. This is obviously related to the initial conditions,
similar to the situation with the prestellar cores.

\section{CONCLUSION}

We have presented a modification of amodel developed earlier for
the calculation of the thermal structure of a collapsing
protostellar cloud. The modified model is intended for the
calculation of the prestellar evolutionary stage of the cloud, as
well as modeling of the accreting envelope after the formation of
the protostar. The main improvements of the model are the
introduction of the sink-cell formalism, calculation of the
luminosity using a model for the evolution of the young star, and
replacement of the Eddington approximation with the diffusion
approximation with a flux limiter in the treatment of the
radiative transfer.

We have applied the model to the evolution of a protostellar
cloud, right up to the formation of the first hydrostatic core and
the subsequent accretion phase. In the inner region of the
accreting envelope ($r<10^{3}{-}10^{4}$~AU), the temperature is
governed by the radiation from the young star. In this zone, the
temperatures of the gas and dust are equal and are well described
by the distribution $T\propto r^{-1/2}$. The thermal structure of
the outer layers of the envelope ($r>10^{3}{-}10^{4}$~AU) is
determined by interstellar radiation; the gas and dust
temperatures rise toward the edge and may differ significantly.

We have used our model to explain the distributions of the surface
density and temperature averaged over the line of sight derived
fromobservations of low mass prestellar and protostellar cores of
molecular clouds obtained with the Herschel Space Observatory. The
model and observed temperature distributions are in good
agreement, for both the prestellar and protostellar cores. At the
same time, the distributions of the surface density display some
differences, probably due to the incorrect assumption that the
cloud was initially in a quasi-hydrostatic state. The model can be
adapted for multi-dimensional calculations, and can also be used
to calculate the chemical structure of protostellar objects.

\bigskip
\bigskip

The authors acknowledge Ralf Launhardt for providing observational
data and B.M.~Shustov for useful discussions.

\end{document}